\pdfoutput=1
\documentclass[lettersize,journal]{IEEEtran}
\usepackage{amsmath,amsfonts}
\usepackage{algorithmic}
\usepackage{array}
\usepackage[caption=false,font=normalsize,labelfont=sf,textfont=sf]{subfig}
\usepackage{textcomp}
\usepackage{stfloats}
\usepackage{url}
\usepackage{verbatim}
\usepackage{graphicx}

\usepackage{multirow}
\usepackage[table,xcdraw]{xcolor}
\usepackage{colortbl} 
\usepackage{tikz}
\usepackage{amsfonts}
\usepackage{amsmath}
\usepackage{filecontents}
\usepackage{booktabs}
\usepackage{dsfont}
\usepackage{nicematrix}
\usepackage{graphicx}
\usepackage{float}
\usepackage{lscape} 
\usepackage[activate={true,nocompatibility},final,tracking=true,kerning=true,spacing=true,factor=1100,stretch=10,shrink=10]{microtype}

\usepackage{balance}
\begin{document}

\title{Transparent Malware Detection With Granular Assembly Flow Explainability via  Graph Neural Networks}

\author{
    \IEEEauthorblockN{Griffin Higgins, Roozbeh Razavi-Far, \textit{Senior Member, IEEE}, Hossein Shokouhinejad, and Ali A. Ghorbani, \textit{Senior Member, IEEE}}
    \thanks{G. Higgins, R. Razavi-Far, H. Shokouhinejad, and A. Ghorbani are with Canadian Institute for Cybersecurity, University of New Brunswick, Fredericton, Canada (e-mail: griffin.higgins@unb.ca).}
}


\maketitle

\begin{abstract}

As malware  continues to  become increasingly  sophisticated, threatening,  and
evasive, malware detection systems must keep pace and become equally
intelligent, powerful, and \textit{transparent}.  In this paper, we propose
Assembly Flow Graph (AFG) to comprehensively represent the assembly flow of a
binary executable as graph data. Importantly, AFG can be used to extract
granular explanations needed to increase transparency for malware  detection
using Graph Neural Networks (GNNs).  However, since AFGs may be large in
practice, we also propose a Meta-Coarsening approach to improve computational
tractability via graph reduction. To evaluate our proposed approach we consider
several novel and existing metrics to quantify the granularity and quality of
explanations. Lastly, we also consider several hyperparameters in our proposed
Meta-Coarsening approach that can be used to control the final explanation size.
We evaluate our proposed approach using the CIC-DGG-2025 dataset. Our results
indicate that our proposed AFG and Meta-Coarsening approach can provide both
increased explainability and inference performance at certain coarsening levels.
However, most importantly, to the best of our knowledge, we are the first to
consider granular explainability in malware detection using GNNs.

\end{abstract}


\section{Introduction}

\IEEEPARstart{M}{alware} is software  explicitly designed with malicious intent
for the purposes of disrupting  digital information and  information systems.
The use of machine learning (ML) and deep learning (DL) to automatically detect
malware has proven extremely promising. This is especially  true for graph
learning-based methods that operate on graph data extracted from program flow
\cite{bilot_survey_2024}. However, since GNNs are not \textit{intrinsically
explainable}, explanation methods must be  used to derive explanations related
to a particular prediction. While different types of explanations are possible,
extracting an explanatory subgraph  related to graph classification (e.g.,
malware detection) is of special interest to improve model transparency. Such  a
subgraph can guide our understanding  of which underlying phenomenon is
influential to a given prediction (i.e., malicious behavior). Subgraph
explanations can  also help confirm that an underlying prediction, irrespective
of correctness, is  being made in a non-spurious manner. Most importantly, the
task of malware detection also demands heightened transparency as the cost of
false negative predictions or misleading explanations are extremely high.

Building on this, it is also important to assess the explanation
\textit{quality} for a given  subgraph. This  can  be  viewed and  measured  in
two fundamental ways. Specifically, an  explanation subgraph can be  either
\textit{sufficient}, where the extracted subgraph matches the prior prediction
of the model on the original sample or \textit{necessary} where the original
sample with the subgraph removed generates  a  different prediction  from  the
prior  models prediction  on  the original sample. However, when evaluated
together, an explanatory subgraph that is able  to  achieve both  a  highly
necessary  and sufficient  explanation gives rise  to  a  further  notion  of  a
highly \textit{characterized} explanation \cite{amara_graphframex_2024}. This is
similar to the way that F1 score provides a deeper understanding of classwise
performance with respect to both precision and recall.

Furthermore, it is important to also consider explanation \textit{granularity}.
While prior  works  have  considered  GNN  malware  explainability  on  CFGs, a
key limitation  is  that  they  are  ultimately  unable to  provide highly
characterized instruction  level  subgraph explanations.  This   is not a
limitation   of  GNNs  or  methods   per  se, but  rather  of  the underlying
CFG  graph  structure.  In  a  CFG,  nodes are represented  as sequences  of
non-branching  assembly  instructions, formally referred  to as basic blocks,
that are  directly connected  to other nodes by edges  that represent  branching
(e.g.,  jump  instructions). However,  since the instructions  do  not  play  an
observable  role  in  the CFG graph structure they  therefore cannot  be
explained  via GNNs. Additionally,  since basic blocks may contain many assembly
instructions it becomes ambiguous as to which instructions contribute to the
explained behavior. By comparison, granular instruction level explanations are
superior and needed since they can provide unambiguous explanations needed to
enhance model transparency. While some related works  do consider aspects of
instruction level program flow via opcode graphs as Markov chains
\cite{anderson_graph-based_2011, runwal_opcode_2012, hashemi_graph_2017,
khalilian_g3md_2018, kakisim_metamorphic_2020, fok_clustering_2021,
kakisim_sequential_2022}, our work  is  distinct and novel since  we consider
the assembly instruction level flow of an entire program as a graph, that  we
term Assembly Flow Graph (AFG). In our work, we use AFG to  improve detection
performance and strongly characterize the predicted  behavior of  a sample
through the use of GNNs. This approach is important because it can provide
granular explanations to help understand and characterize malware behavior. This
is especially useful for assisting malware analysis where the number of
instructions are large.

However, many challenges related  to the large size and scale  of AFGs make most
aspects of applying GNNs  (preprocessing, embedding, training, explaining, etc.)
non-trivial. To address this, we apply coarsening, a graph reduction technique,
to reduce both nodes and edges in the original graph to a prespecified size.
Critically, we select coarsening since it aims to maintain structural similarity
to the original graph while also providing nodewise backtrackability from
reduced to original nodes.

\begin{figure}[t]
    \includegraphics[width=0.5\textwidth]{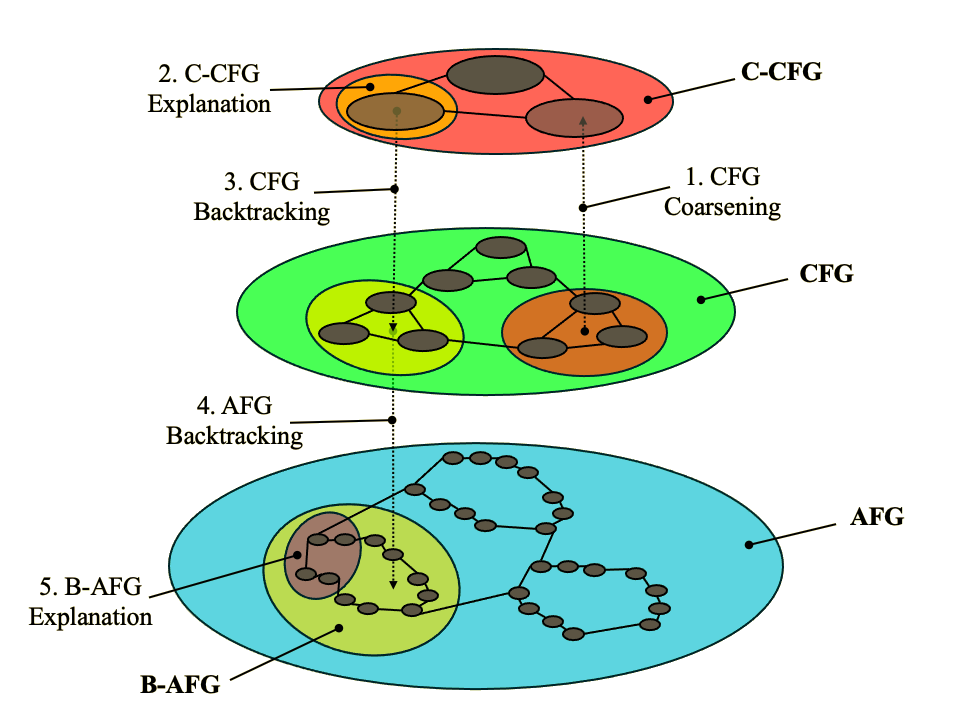}
    \caption{High-level diagram of our proposed approach with training and
    embedding phases omitted. CFG and starting point is shown in the middle
    green patch, coarsened CFG (C-CFG) shown in the top red patch, and AFG is
    shown in the bottom blue patch. The C-CFG explanation and backtracked AFG
    (B-AFG) are shown in yellow. The most important B-AFG explanation subgraph
    is shown in purple.} 
    \label{fig:projection_diagram} 
\end{figure}

Nonetheless, applying coarsening directly on AFG remains computationally
challenging. To alleviate this, we propose Meta-Coarsening, shown at a high
level in Fig. \ref{fig:projection_diagram}, where we first coarsen the CFG, (i.e.,
AFG meta-structure) to approximate coarsening AFG directly. Once a coarsened
C-CFG, dataset is constructed a GNN model is then trained and an explainer is
used to derive explanations. Here, the most important C-CFG explanation subgraph
is then backtracked onto the underlying AFG. Afterwords, an entirely separate
backtracked AFG (B-AFG) dataset is used to train a separate GNN model. An
explainer is then used to derive the final B-AFG subgraph level explanation
\textit{at the assembly instruction level}. This approach has the advantage that
any subgraph of the AFG  can be arbitrarily examined under a varying level of
coarsening and adjusted to fit the computational requirements or explanation
expressiveness. 

To evaluate our proposed approach, we conduct experiments on the Dynamically
Generated Graphs for  Malware Analysis (CIC-DGG-2025)
\cite{shokouhinejad_consistency_2025}. Specifically, we consider several
coarsening methods and coefficients and evaluate our work using the strongest
available explainability metrics. We also propose several novel metrics to
assess quality and granularity of explanations. Importantly, to the best of our
knowledge, we are the first to propose granular explainability at the assembly
level for malware detection.

In summary, our contributions are as follows:

\begin{itemize}

    \item We propose a novel program graph type, AFG, to represent the assembly instruction flow of a binary executable.

    \item We capture granular explainability of AFG towards enhanced GNN transparency.

    \item We propose a novel Meta-Coarsening approach to improve computational tractability over large graphs.

    \item We define several novel explanation metrics to quantify explanation granularity and quality.

    \item We consider several hyperparameters to control the size of explanations.

\end{itemize}

The remainder of this paper is organized as follows. In Section
\ref{sec:literature_review} we conduct a literature review. In Section
\ref{sec:background} we provide background. Section \ref{sec:methodology} we
cover our methodology. In Section \ref{sec:evaluation} we present our
evaluation. In Section \ref{sec:discussion_and_limitations} we discuss the work
and note limitations. Lastly, in Section \ref{sec:conclusion}, we conclude our
work.

\section{Literature Review}
\label{sec:literature_review}

In  this  section,  various  related  works  that  consider  either  graph based
instruction level malware detection and classification (i.e., excluding CFG
level)  or contribute to  explainable graph based malware detection methods
generally are considered.

\subsection{Opcode graph mining for malware detection and classification.}

In \cite{anderson_graph-based_2011},  the authors  propose using  support vector
machines  (SVMs)  to  detect  and  classify malware.  Initially,  all  pairs  of
successive  instructions  are recorded  with  observed  counts in  matrix  form.
Afterwards,  the  matrix is  normalized  to  form  a  Markov chain  where  nodes
represent  instructions  and  edges represent  transition  probabilities.  Graph
kernels are then applied to score the graph and supplied as input to SVMs.

In \cite{runwal_opcode_2012}, the authors propose  the same initial opcode graph
construction approach as \cite{anderson_graph-based_2011} but compare the graphs
directly instead of  applying graph kernels and SVMs.  Specifically, the authors
propose  a score  to  measure the  similarity between  two  Markov chain  opcode
graph  representations. The score is used to compute similarity between known
metamorphic samples to detect matches. The authors set a score threshold to
classify a given sample as malicious or benign.

In       \cite{hashemi_graph_2017},       opcode        graphs       as       in
\cite{anderson_graph-based_2011}, are  used  and power  iteration  is used  to
derive  graph embeddings  input to  an  ensemble of classifiers  to detect
malware. Models  such  as k-nearest  neighbors  (KNN),  Adaboost,  SVM  are used
in  the ensemble.

In \cite{khalilian_g3md_2018},  Graph Mining  for Metamorphic  Malware Detection
(G3MD)   is  proposed   as   an  approach   to   classify  metamorphic   malware
families.  The  approach adopts  the  graph  construction strategy  proposed  by
\cite{runwal_opcode_2012,  anderson_graph-based_2011}   and  attempts   to  find
frequent subgraphs within malicious metamorphic  opcode graphs via graph mining.
The subgraphs  are then saved  as micro-signatures  and matched against  a given
sample. Matches are then vectorized and a classifier is used to classify a given
sample as belonging to a given metamorphic malware family.

In  \cite{kakisim_metamorphic_2020},  the  authors propose  Higher-level  Engine
Signature  based  Metamorphic  Malware  Identification  (HLES-MMI)  to  identify
metamorphic malware.  The approach uses  a co-opcode graph constructed  for each
malware family as  in \cite{runwal_opcode_2012}. The authors  also propose Graph
Similarity based Metamorphic Malware  Identification method (CGS-MMI) to compare
a sample with all co-opcode graphs for each family. Furthermore, the size of the
graphs are reduced as to their largest connected component.

In  \cite{fok_clustering_2021},  opcode  sequences   are  first  extracted  from
disassembled  malicious  families and  constructed  as  opcode graphs  following
\cite{runwal_opcode_2012}, \cite{khalilian_g3md_2018}.  Opcode filtering  for low
frequency  opcodes is  also applied  to  reduce the  size of  the opcode  graph.
Afterwards, similarity  comparison as  proposed in  \cite{runwal_opcode_2012} is
performed  to  group malware  families.  Sub-family  clusters are  then  derived
using multi  round Density-Based Spatial  Clustering of Applications  with Noise
(DBSCAN).  Afterwards, representative  sub-family opcode  graphs signatures  are
then used to detect and classify samples that may be polymorphic or metamorphic,
or both.

In \cite{kakisim_sequential_2022},  a Sequential Opcode  Embedding-based Malware
Detection (SOEMD)  approach is  proposed that aims  to capture  opcode sequences
through weighted  random walks.  However, only strongly  connected substructures
are considered via  an edge selection process. Afterwards, a  Skip-Gram model is
used to embed  all walks input as unique one-hot  encodings. The authors justify
the approach to avoid having long opcode sequences.

\subsection{Explainable GNN for malware detection and classification.}

In \cite{herath_cfgexplainer_2022},  a CFGExplainer  is proposed to  explain
GNN-based malware  detection at the  CFG level. CFGExplainer captures  important
CFG subgraphs  that  relate  to  the  prediction of  the  GNN  model.
Specifically, malicious attribute CFGs are used to train a GNN model where node
attributes are constructed via  assembly instruction  information to  produce
initial  CFG node embeddings. Furthermore, a two-stage approach  is used.
Firstly, node embeddings are  learned by  a  model  that assigns  node
embedding  scores. Secondly,  the assigned  learned scores  are used  to obtain
subgraph importance.  Several GNN explainers such as  GNNExplainer, PGExplainer,
and SubgraphX  are also evaluated against the proposed CFGExplainer.

In   \cite{shokouhinejad_consistency_2025},  several   GNN  explainers,   namely
GNNExplainer, PGExplainer,  and CaptumExplainer  are considered to  evaluate the
explainability  of  GNN-based  malware  detection.  In  this  work,  dynamically
generated CFGs  are extracted from benign  and malicious samples. Nodes  in each
sample are embedded  using assembly instructions before being  passed through an
auto encoder. Embedded samples are then  trained and tested using a standard GCN
model. Samples are then explained using the provided explainers and evaluated. A
novel  explanation  selection method,  Greedy  Edge-wise  Composition (GEC),  is
proposed  in  place of  Top-Edge  Selection  (TES)  to  address the  problem  of
having  many disconnected  subgraphs at  various levels  of selection  sparsity.
Furthermore, a novel explainer, RankFusion, is proposed to aggregate and improve
explanations from multiple high performing explainers.

In \cite{shokouhinejad_dual_2025},  a dual  explainability framework  is
proposed that considers GNN-based malware  detection subgraph explanations.
Specifically, as  in  \cite{shokouhinejad_consistency_2025},  dynamically
generated  CFGs  are extracted from  benign and  malicious samples and embedded.
Several  GNN models such as GCN, GAT, and GraphSAGE  are then trained and
tested. Furthermore, three GNN explainers, Integrated Gradients,  Guided
Backpropagation, and Saliency, are used to extract important subgraphs
signatures from benign and malicious samples and build  a verified query box.
When evaluating a unseen sample signatures in the  query box are tested for
subgraph isomorphic or  monomorphic matches within the unseen sample graph
through an approach termed SubMatch. 

In  \cite{shokouhinejad_recent_2025}, a  survey  of recent  advances in  malware
detection  and  explainability  with  GNNs   is  provided.  Emphasis  is  placed
on  malware  analysis   and  datasets,  graph  reduction   strategies,  and  GNN
explainability are covered.

\section{Background}
\label{sec:background}

Here we cover background needed to understand different aspects of our work.
Specifically, we provide information on our sample focus area, Windows PE files,
as well as CFGs, coarsening, GNNs, and GNN explainability.

\subsection{Windows PE}
\label{subsec:windows_pe}

The Portable  Executable (PE)  file format  is a  specification used  by Windows
operating systems to  describe the architecture of executable  files, and object
files referred  to as PE and  Common Object File Format  (COFF), respectively.
Of course,  the executable  is  portable,  meaning it  can  run  on many
different hardware architectures,  such as x86-32.  The PE executable format  is
organized with  the MS-DOS  Header, MS-DOS  Stub, PE  Header, Section  Headers,
and  Image Pages. Many other subsections (e.g., .data  and .text) are equally
important for the functioning of the format. For a  comprehensive review of the
PE file format please see  \cite{karl-bridge-microsoft_pe_nodate}. Most
importantly,  since the PE  format exists  as  part of  (currently) the  most
dominant consumer  facing operating system, i.e., Windows. As such, it is
entirely reasonable to expect an adversary to target the PE format in order to
indirectly perpetrate harm against consumers generally.

\subsection{Control Flow Graphs}
\label{subsec:cfg}

CFGs are  graph data representing an  abstracted view of program  by considering
nodes as sequences  of non-branching assembly instructions.  Therefore, any type
of branching can be represented by  an edge to another basic block. Importantly,
nodes can  contain many  types of  data related  to it  apart from  the assembly
instructions themselves.

\subsection{Graph Coarsening}
\label{subsec:gc}

Graph  coarsening is  one  of several  existing categories of graph reduction
techniques, which aim to  reduce the  size of  a  graph via  the  contraction of
disjoint  sets  of  connected nodes  into supernodes. Importantly, during
contraction the  overall graph retains a similar representative  topology  or
structure  as   the  original.  This  is  distinct from  other  related  graph
reduction techniques  such  as  sparsification  and condensation,  since
information is  not  necessarily  synthesized or  removed. Instead,  all
information  is  still  entirely  backward  referential  since  a surjective
mapping between  nodes  in  the original  and  coarsened levels  are maintained.
This means that given a  coarsened graph it is possible to backtrack supernodes,
and hence subgraphs, to the  original graph, so long as the original graph
structure is known so an edgewise subgraph can be induced. Lastly, coarsening
methods also allow for coarsening to a specific size as specified by the $r$
coarsening coefficient. Specifically, the coefficient reduces a graph to $1 - r$
of its original size.

\subsection{Graph Neural Networks}  
\label{subsec:gnn}

GNNs  operate  on attribute  graph-structured  data  to learn  high  dimensional
representations for  graph primitives towards  a specific task. Here,  tasks can
range in  level, from  nodes and  edges to  subgraphs and  graphs. Additionally,
tasks  may  be supervised  (e.g.,  classification  or regression),  unsupervised
(e.g., clustering), or semi-supervised.

Formally,  the   input  GNN   takes  the   form  $\mathcal{G}   =  (\mathcal{V},
\mathcal{E})$, where $\mathcal{E}$ represents the edge list (or adjacency matrix
similarly) and $\mathcal{V}$ represents the set of nodes or vertices. We opt for
the  former  term.  Importantly,  in  an  attribute  graph  nodes  are  assigned
attribute vectors,  i.e., message initializations, stored  in $\text{\textbf{X}}
\in \mathbb{R}^{n \times d}$ of  the form $\text{\textbf{x}}_v \in
\mathbb{R}^d$, where $n$ is the number of nodes, $d$ is the attribute vector
dimension, and $v$ is a node in $\mathcal{V}$. These  initialized message
vectors are then updated, and learned  by GNNs  during training, via  message
passing  between neighboring nodes. The  exact neighborhood  is defined  thought
the  function $\mathcal{N}$. Additionally,  $l$ layers,  or rounds,  of message
passing are  performed where neural  networks  optimize  the  final  attribute
vectors  for  each  node.  The generalization of this process takes the form:

\begin{equation}
\label{eq:message_passing}
\textbf{x}_v^{(l)} = \gamma^{(l)}\left( \textbf{x}_v^{(l-1)}, \bigoplus_{u \in \mathcal{N}(v)} \phi^{(l)}(\textbf{x}_{v}^{(l-1)}, \textbf{x}_{u}^{(l-1)}) \right)
\end{equation}

Here, $\bigoplus$  is a permutation  invariant aggregation function  (e.g., sum,
mean, or  max). Furthermore,  $\phi$ and  $\gamma$ are  differentiable functions
used  to update  messages. Edges  may also  optionally contribute  to a  message
update with separate input information, however we do not consider this here.

\begin{figure*}[ht]
    \includegraphics[width=1\textwidth]{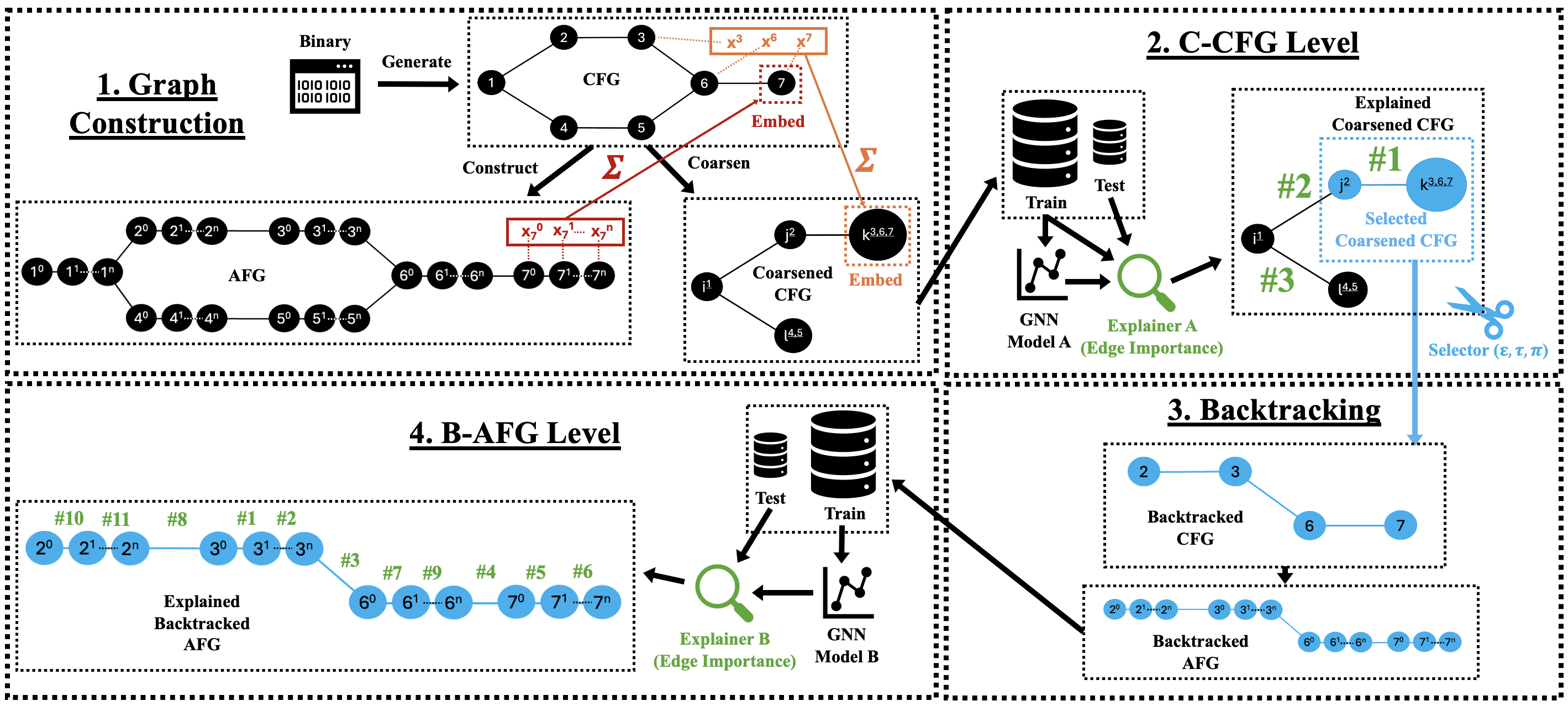}
    \caption{Detailed diagram showing four phases of processing.}
    \label{fig:system_diagram}
\end{figure*}

\subsection{GNN Explainability}
\label{subsec:gnn_ex}

GNN  explainability   methods  can  fall   into  two  categories:   factual  and
counterfactual. In factual methods, the target  is to identify the components of
the  input  graph  that  directly  affect  the  GNN  model's  prediction.  These
components  comprise information  about  nodes, edges,  and  node features  that
GNN  models use  during  message  passing and  prediction.  On  the other  hand,
Counterfactual explanation  methods pinpoint how  a minimal perturbation  of the
input graph could  change the model prediction and describe  how sensitive it is
to graph structure or to node features.

GNN explainers can  be divided into post-hoc  and self-interpretable approaches,
depending  on  their  architecture.  Post-hoc explainers  leverage  the  trained
model  to produce  explanations  either via  black-box  or white-box  techniques
while   self-interpretable  explainers   integrate  the   explanation  mechanism
within  the GNN  model  design. Indeed,  self-interpretable  explainers rely  on
information-theoretic  or  structural  constraints  to  highlight  an  important
subgraph used for both prediction and explanation tasks.

Also, explanations  can be classified into  two main categories in  terms of the
scope of  analysis: instance-level and model-level  explanations. Instance-level
explainers  generate  interpretations  based  on  a  specific  input  graph  and
provide localized  insights into individual predictions.  Model-level explainers
aim  to  determine  the  overall  decision-making behavior  of  the  GNN  model,
seeking  broader  patterns  that  likely  hold across  the  majority  of  inputs
\cite{shokouhinejad_recent_2025}.

\section{Methodology}
\label{sec:methodology}

Here, we present our proposed Meta-Coarsening approach covering all phases as
shown in Fig. \ref{fig:system_diagram}. In Section
\ref{sec:graph_construction} we first describe phase 1, the initial graph
construction process in for CFG, AFG, and C-CFG graphs. Importantly, while the
CFG generation process must proceed the AFG and C-CFG construction that latter
two graph can be constructed independently from the CFG. Additionally, we also
describe the instruction encoding process used in the work. Next, in Section
\ref{sec:c_cfg_level}, we consider the operations related to the C-CFG level of
our proposed approach. Specifically, we cover the training and explaining of
C-CFGs as well as the selection of explanation needed for the backtracking
stage. In Section \ref{sec:backtracking}, we describe the backtracking process
to reach the B-AFG and the difference in embedding used as compared to C-CFG.
Lastly, In Section \ref{sec:b_afg_level}, we finish with a brief description of
the training and explanation process to reach the final instruction level
explanation.

\subsection{Graph Construction}
\label{sec:graph_construction}

\subsubsection{CFG Generation} 
\begin{figure}[h]
    \centering
    \includegraphics[width=0.40\textwidth]{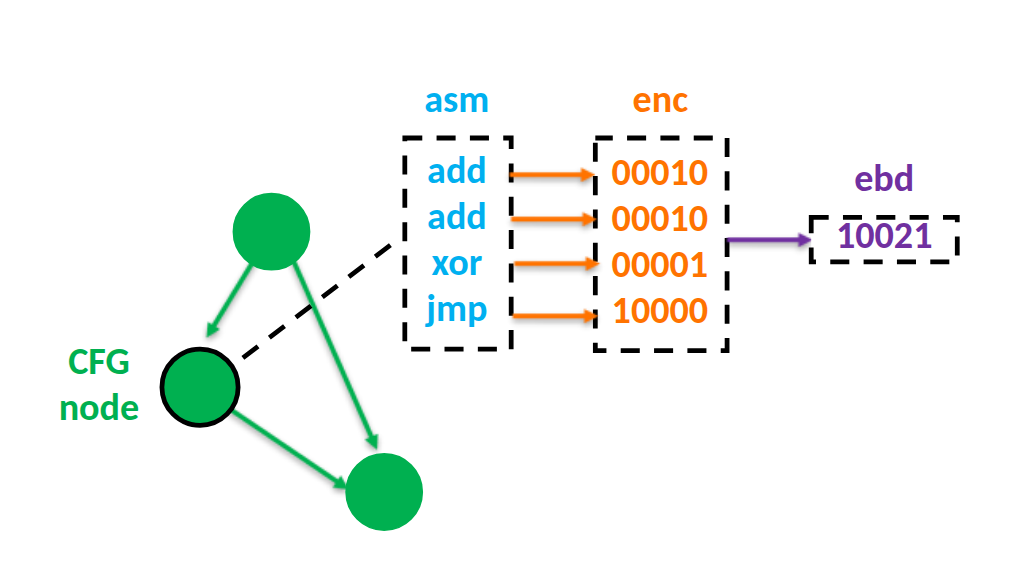}
    \caption{CFG embedding procedure.}
    \label{fig:cfg_graph_embedding}
\end{figure}

We begin by first obtaining attribute CFGs of
benign and malicious Windows  PE x86 executable samples. Here, we consider CIC
Dynamically Generated Graphs for Malware Analysis (CIC-DGG-2025)
\cite{shokouhinejad_consistency_2025}, discussed further in Experimental Setup.
angr  Python library (version 9.2.89) to dynamically generate CFGs.  angr
leverages symbolic execution  and constraint solving  to recover program  flow
and   CFGs \cite{shoshitaishvili2016state, stephens2016driller,
shoshitaishvili2015firmalice}.  angr also  relies on various libraries  such as
capstone for instruction encoding that we  use to derive instruction features in
our work.  The resulting CFG can be used to access the assembly instructions,
and other information, located within each CFG node to construct the AFG,
discussed hereafter. Lastly, the embeddings for CFG nodes are simply the sum of
all instruction encoding that belong to them, as shown in Fig.
\ref{fig:cfg_graph_embedding}. 

\subsubsection{AFG Construction} 
\begin{figure}[h]
     \centering
     \includegraphics[width=0.43\textwidth]{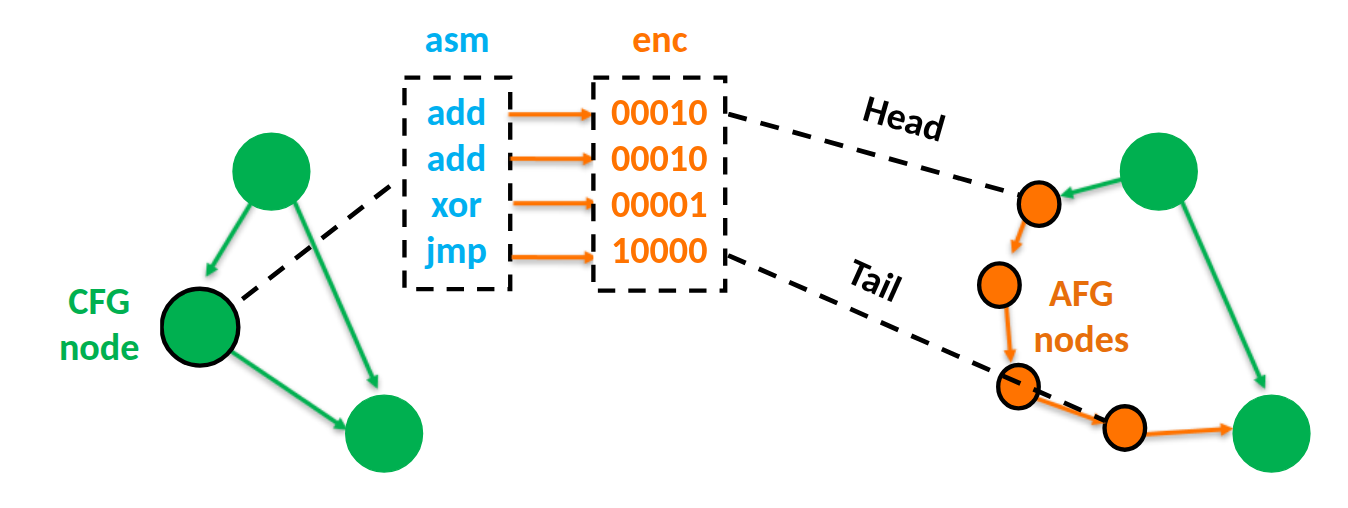}
     \caption{AFG construction and embedding procedure for one node.}
    \label{fig:afg_graph_construction}
\end{figure}

Given  a CFG, captured dynamically from  a
Windows PE x86 executable  binary, as  is the  case for CIC-DGG-2025
\cite{shokouhinejad_consistency_2025},  we construct  an AFG, as shown in Fig.
\ref{fig:afg_graph_construction}, by  substituting all CFG  nodes, i.e.,  basic
blocks, with their  respective non-branching sequential  assembly  instructions.
Such instructions  can  be represented  as a  singly linked  list,  that  we
refer to hereafter  as  the \textit{instruction list} for a given  CFG node.
Importantly, all incoming edges to the original CFG node are  reassigned to the
respective instruction list head node and all outgoing edges from  tail node. 

\subsubsection{CFG  Coarsening}  Due  to  the  large size  of  many  AFGs,  the
direct application of coarsening often becomes computationally intractable or
extremely costly  in  practice. To address  this, we  propose  to instead
coarsen  the  CFG  which is computationally  tractable to  produce   a  C-CFG.
In  our   work,  we  utilize the   coarsening  library from
\cite{loukas_graph_2018}.  Specifically,  we aim to  take  the advantage  of the
hierarchical  relationship  between   the various  levels  of  granularity and
backtrack   explanations  from   C-CFG supernodes   to  CFG   basic  block nodes
and   eventually  to   AFG instructions  nodes. We  argue   that coarsening  the
CFG,  to approximate coarsening the underlying AFG is reasonable. This is
partially owing to the  computational tractability of coarsening a CFG,  but
mainly due  to the \textit{guaranteed} structural similarity between CFG and
AFG. By this, we simply mean that that nodes in CFG with specific in and out
degrees are guaranteed to retain their respective in and out  degrees in the
AFG, just at opposite ends of the substituted instruction list. Otherwise, apart
from the increase in nodes the general structure of AFG remains well constrained
by its respective CFG. Moreover, it is also guaranteed that only  path graphs or
linked  lists will be substituted in place  of CFG nodes rather than highly
complex and unpredictable graph structures. To compute the embedding  of each
C-CFG  supernode we simply sum  the embeddings  of all contracted CFG  nodes,
whose  embeddings are  the sum  of all instruction encodings  that belong  to
them. 


\begin{table}[h]
    \centering
    \caption{Instruction features and dimension for node embedding.}
    \label{tab:feature_table}
    \begin{tabular}{ll c ll}
        \toprule
        \textbf{Feature} & \textbf{Dim} & & \textbf{Feature} & \textbf{Dim} \\
        \midrule
        $\text{prefix}_0$    & 2   & & xop cc            & 5   \\
        $\text{prefix}_1$    & 4   & & sse cc            & 21  \\
        $\text{prefix}_2$    & 7   & & avx cc            & 1   \\
        $\text{prefix}_3$    & 2   & & avx sae           & 1   \\
        $\text{opcode}_0$    & 200 & & avx rm            & 5   \\
        $\text{opcode}_1$    & 191 & & eflags            & 1   \\
        $\text{opcode}_2$    & 37  & & fpu flags         & 1   \\
        $\text{opcode}_3$    & 1   & & modrm offset      & 10  \\
        rex                  & 17  & & disp offset       & 12  \\
        addr size            & 3   & & disp size         & 5   \\
        modrm                & 256 & & imm offset        & 10  \\
        sib                  & 255 & & imm size          & 5   \\
        sib scale            & 21  & &                   &     \\
        \bottomrule
    \end{tabular}
\end{table}

\subsubsection{Instruction Encoding}  For encoding instructions we consider the
most representative and variable categorical  feature subset, shown  in Table
\ref{tab:feature_table}, of  a given  instruction node, generated by the
capstone library \cite{qemu_capstoneincludecapstonex86h_nodate} angr
\cite{shoshitaishvili2016state, stephens2016driller,
shoshitaishvili2015firmalice}. Additionally, we  consider the prior  frequency
of all well  defined categorical features across the millions of instructions in
our training  set (derived from hundreds of samples across independent
datasets). Specifically, we observe that while many  one-hot encoding categories
are reserved  many are not  observed in training. Thus, we can save one-hot
encoding space by only considering observed categories and remapping them  to a
reduced one-hot  space. However,  for each category,  we include  an  additional
bit for  each  feature  to  capture  the unlikely occurrence  of  out of
vocabulary  instruction  level categories for the reduced  one hot  mapping with
respect  to our  dataset. In total, we consider a 1076 dimensional vector with
additional 27 bits of out of vocabulary space, bringing  the  total  to  1103
bits.

\subsection{C-CFG Level}
\label{sec:c_cfg_level}

\subsubsection{(C-)CFG Training and  Explaining} During the training phase  a
GNN model is selected  and trained on the  embedded (C-)CFG training graphs  and
tested on the test  graphs. Its  weights are  then saved  for future  use with
respect to explanation and  further evaluation.  Once training is completed an
explainer is selected and each test sample is explained using the GNN model,
prediction, and test sample as input. The returned output from the explainer are
the corresponding set of edge importance weights (C-)CFG.

\subsubsection{(C-)CFG Selection}  After a given explanation is provided for a
specific (C-)CFG sample, a subgraph must be selected according to a  given
selection algorithm $\tau$ e.g., Top Edge Selection, and selection policy $\pi$,
e.g., keep 5 percent of the top nodes. Together, a final subgraph explanation
can be provided at the (C-)CFG level and used in the next phase for backtracking
to the AFG.

\subsection{Backtracking}
\label{sec:backtracking}

\subsubsection{(B-)AFG Construction}

Using the prior C-CFG subgraph explanation a AFG subgraph can be backtracked and
defined throught the mapping between supernodes in the (C-)CFG, basic block node
in the CFG, and instruction nodes in the AFG. Once the B-AFG subgraph is defined
embeddings for each node are computed. However, unlike  the  prior embedding  of
the (C-)CFG,  (B-)AFG embedding cannot rely on an aggregated embedding
representation to  reduce  its  memory  footprint. Instead,  each  node  exists
in  its  pure unaggregated form.  This is an attractive  property for
explainability but raises computational concerns from a space  perspective since
AFGs may be  very large compared  to CFGs. Therefore, it is worthwhile to
consider our  prior feature selection for  (C-)CFGs and  opt for a  reduced
feature  or feature  subset. Here,  we  only  consider  the  single reduced
$\text{opcode}_0$  feature of 201 dimensions including  the one  out of
vocabulary  bit. This feature is specifically selected since it can capture the
operations being performed at the assembly level while enabling backtracking
over larger B-AFG subgraphs needed to obtain improved $\text{Fidelity}{+}$
score, discussed in Section \ref{sec:metrics}.

\subsection{B-AFG Level}
\label{sec:b_afg_level}

\subsubsection{B-AFG Training and Explaining}  After all B-AFGs are constructed
and embeddings are computed,  a completely new GNN model is  used and  trained
on  the B-AFG graphs.  Additionally, only after the (C-)CFG model is fully
trained and B-AFG graph are defined  can the B-AFG model begin  training. After
training  is concluded the  test  samples can then  be explained using an
explainer. Lastly, granular explanations at the instruction level can be
provided and subsequently assessed.

\section{Evaluation}
\label{sec:evaluation}

In  our work,  we evaluate  several relationships  between \textit{coarsening},
\textit{inference}, and  \textit{explainability} as applied to our novel
Meta-Coarsening approach.  Specifically, we formulate the following research
questions:

\begin{itemize}

  \item \textbf{RQ1:}  Is the application of \textit{coarsening} harmful or
  helpful for improving \textit{inference} performance with respect to
  coarsening method and coefficient on average accuracy?

  \item   \textbf{RQ2:}  Is the application of \textit{coarsening} harmful or
  helpful for improving \textit{explainability} with respect to coarsening
  method and coefficient on $\lambda$ score (introduced in Section
  \ref{sec:metrics})?

  \item \textbf{RQ3:}  Are there  any noticeable \textit{classwise differences
  or trends} with respect to  \textit{inference} or \textit{explainability}?

\end{itemize}

\subsection{Experimental Setup}

\subsubsection{Dataset}    In    this    work,   we    consider    the
CIC-DGG-2025 \cite{shokouhinejad_consistency_2025}   dataset   that  provides
our   initial attribute  CFGs across  two datasets, namely, DikeDataset
\cite{dikedataset} and PE Malware Machine Learning Dataset (PMMLD)
\cite{practicalsecurity2024pe}. Here, benign samples belong exclusively to the
DikeDataset and malicious samples belong exclusively to PMMLD dataset.

\begin{figure}[h]
    \centering
    \subfloat[DikeDataset sample sizes in nodes and edges (benign).]{
        \includegraphics[width=0.98\columnwidth]{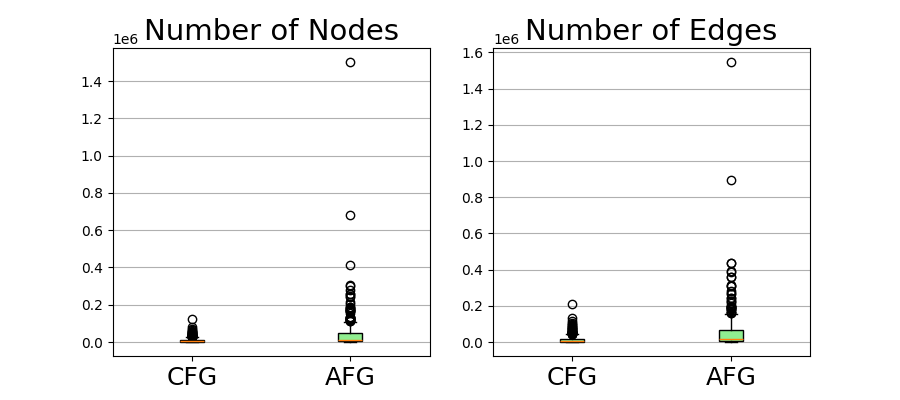}
        \label{fig:box_dike}
    }
    \hfil 
    \subfloat[PE Malware Machine Learning Dataset sample sizes in nodes and edges (malicious).]{
        \includegraphics[width=0.98\columnwidth]{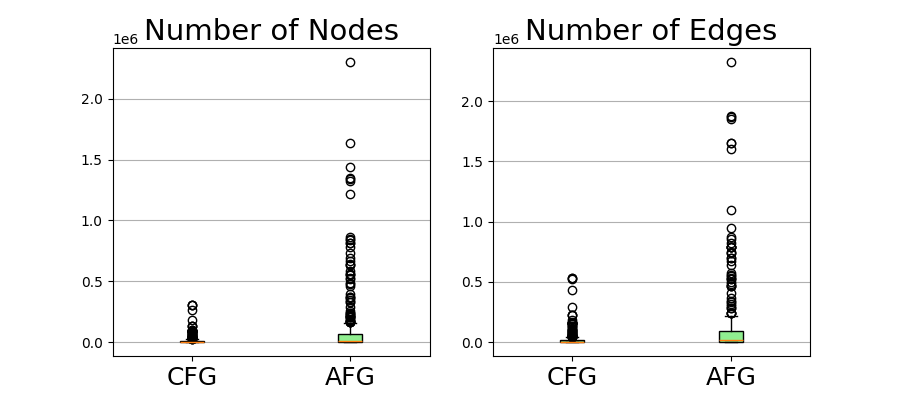}
        \label{fig:box_pmml}
    }
    \caption{Size comparison of datasets with respect to CFG and AFG number of nodes and edges.}
    \label{fig:box}
\end{figure}

Furthermore, we consider a stratified classwise, but not dataset wise, train
test split of 80\% between all non-isomorphic CFGs samples in the dataset with
45\% benign and 55\% malicious samples.  We fix the  training and  test sets
for all  models and  experiments in  our work. Additionally,  the frequency  of
all well-defined  categorical features  (e.g., opcode) are recorded for each
training sample to construct the instruction encoding discussed prior. Next, we
proceed to construct the full AFG for each remaining CFG. The size, in  number
of nodes and  edges, for each dataset  is shown  in  Fig. \ref{fig:box}.
Importantly, a  very  large difference between CFGs and AFGs can be observed
helping to motivate the application of coarsening due to the large size
difference between nodes and edges in CFGs and AFGs in each dataset.

\subsubsection{GNN  Model  and  Hyperparameters}  In  our  work,  we  train  18
Graph Convolutional Network (GCN)  \cite{kipf_semi-supervised_2017} models. All
models are provided  with the same  training configuration, namely, three
layers, 50\% dropout, input dimension equal to the length of the (C-)CFG/B-AFG
node attribute vector,  hidden  dimension of  128,  output  dimension  of 2.
Furthermore,  for the optimizer  we utilize  Adam with  a learning  rate of
0.001. In  terms of  loss function, we  use the Cross Entropy  (CE) loss.
Lastly, we  train our models  for 150 epochs with a fixed batch size of 8. After
training is complete, we consider the model with the lowest validation loss for
evaluation.

\subsubsection*{GNN Explanation and Hyperparameter Selection} In our work, we
consider the Integrated Gradients (IG) explainer provided by the Captum library
\cite{Captum}. We also provide additional explainability comparison with other
captum explainers in Appendix \ref{charact_analysis}. However, while other
explainers, such as GNNExplainer \cite{GNNExplainer}  and PGExplainer
\cite{PGExplainer} are capable of  node level  explanations, other works have
observed  that such explanations are  not as  consistent or  high quality as
the former in a similar setting  to ours \cite{shokouhinejad_consistency_2025}
and are not considered further. Additionally, we define $\pi$, i.e., selection
stopping policy, as the selection of ranked edges, output from a given selection
method $\tau$,  up to a percentage $\epsilon$ of  nodes in the original graph,
selected as  a edgewise subgraph of the original  graph. If the percentage is
less than the number of nodes in the graph we simply select  the maximum of
either the logarithmic number of nodes in the graph or one edge. This guarantees
that a non  empty graph  is always selected.  Furthermore, we define the
unimportant subgraph  as the  edgewise subgraph  of  the original graph, for all
non-important edges. We fix our chosen $\epsilon$ value at 0.10 for all
experiments. Furthermore, we choose TES as our $\tau$ selection method.

\subsubsection{Coarseners}  In  our   work,  we  select  the   Kron  reduction
method \cite{shuman_multiscale_2016}   that  adapts   a  Laplacian   pyramid
transform for  graph  signals  using  downsampling,  reduction,  spectral
filtering,  and interpolation operations on  graphs to approximate the
underlying graphs. While technically  not a  coarsening method
\cite{loukas_graph_2018}, it  is able  to reduce  the  graph  by  a  factor  of
two  at  each  level  and  maintain  the effective  resistance  distance. While
Kron  has  many  drawbacks  it  enables certain  theoretical guarantees that
make  it popular  as a  coarsening method. Additionally,  we  also   select the
edge  variant  of   the  local  variation coarsening  method (Variation Edges)
\cite{loukas_graph_2018}  that relies  on restricted spectral approximation  and
greedy algorithms that  operated on small well-connected sets and select those
with minimal local variation. The approach achieves improved performance over
coarsening methods  such as  Kron. However, other coarsening methods are
available  such as heavy edge,  neighborhood local variation, algebraic
distance, and  affinity. Thought, we do not  consider them simply to limit the
large scope of our existing experimental setting.

\subsubsection{Hardware and  Software} In our  work, we  use Intel Xeon
Platinum 8253 (32) @ 3.000GHz and 128GB of memory  with NVIDIA Quadro RTX
6000/8000 as well as a  Xeon  Silver  4114 (20)  @  3.000GHz  and  64GB  of
memory  with  two  NVIDIA TITAN  V's.  Furthermore,  we utilize Python  version
3.12.9  and  extensively utilize  the  PyTorch Geometric  graph  representation
deep  learning  library \cite{Fey_Lenssen_2019, Fey_etal_2025},  Networkx
\cite{scipyproceedings_11}, as well as  angr and its  modules such as  capstone
\cite{shoshitaishvili2016state, stephens2016driller,
shoshitaishvili2015firmalice}.

\subsection{Metrics}
\label{sec:metrics}

\subsubsection{Inference Metrics} 

In  terms of evaluation metrics, it  is important to separate the  (C-)CFG and
B-AFG  inference and explainability performance since they represent  entirely
separate  models. However,  the B-AFG model is dependent on the (C-)CFG model
and chosen explainer module (\textit{explainer, $\tau$},  and  \textit{$\pi$})
for  defining its  training and  testing  B-AFG datasets. Additionally, since
different $r$ values can produce different (C-)CFG and  B-AFG  sizes they  must
also  be  considered. We use the standard accuracy, precision, recall, and  F1
score.  Specifically, TP is  true  positive, TN is true negative, FP is false
positive, and FN is false negative, formulated below as:

\begin{equation}
\text{Accuracy} = \frac{TP + TN}{TP + TN + FP + FN}
\end{equation}

\begin{equation}
\text{Precision} = \frac{TP}{TP + FP}
\end{equation}

\begin{equation}
\text{Recall} = \frac{TP}{TP + FN}
\end{equation}

\begin{equation}
  \text{F1 score} = \frac{2 \cdot \text{Precision} \cdot \text{Recall}}{\text{Precision} + \text{Recall}}
\end{equation}

\subsubsection{Explainability Metrics} To assess explainability performance, we
use the standard $\text{Fidelity}{+}$  ($fid_{+}$) and  $\text{Fidelity}{-}$
($fid_{-}$) scores where  higher $fid_{+}$ and lower  $fid_{-}$ scores are
preferred, respectively. 

\begin{equation}
fid_{+} = 1 - \frac{1}{N}\sum_{i=1}^{N} \mathds{1} ( \hat{y}_{i}^{G_{C \backslash S}} = \hat{y}_{i} )
\end{equation}

\begin{equation}
fid_{-} = 1 - \frac{1}{N}\sum_{i=1}^{N} \mathds{1} ( \hat{y}_{i}^{G_{S}} = \hat{y}_{i} )
\end{equation}

\begin{equation}
\mathds{1}(x) =
\begin{cases}
1 & \text{if } x > 0 \\
0 & \text{if } x = 0 \\
-1 & \text{if } x < 0
\end{cases}
\end{equation}

Here,  $\hat{y}_{i}$   is  the prior  prediction  on   a  given   sample  $s$,
$\hat{y}_{i}^{G_{C  \backslash S}}$  is  the prediction  of  $s$  with  the
important subgraph removed,  and $\hat{y}_{i}^{G_{S}}$ is the  prediction of $s$
considering only the important subgraph. Since  our task is classification and
not regression, we express  the fidelity scores using  indicator  functions
$\mathds{1}$, also  referred to as accuracy, ($fid_{+/-}^{acc}$) over  predicted
probabilities ($fid_{+/-}^{prob}$) as mentioned in
\cite{amara_graphframex_2024}. 

\subsubsection{Lambda Score} The  characterization  score  ($charact$)  is
the  weighted  harmonic  mean  of $fid_{+}$ and $fid_{-}$, i.e., same as
Micro-F1 for precision and recall, where $w_{+}$ and $w_{-}$ $\in [0, 1]$ and
where $w_{+} + w_{-} = 1$. We assign $w_{+} = w_{-} = 0.5$ as shown below:

\begin{equation}
charact = \frac{w_{+} + w_{-}}{\frac{w_{+}}{fid_{+}} + \frac{w_{-}}{1 - fid_{-}}}
\end{equation}

Furthermore,  we  recognize that  a  ``good"  characterization  score to  be
0.6 based  on  different   experimental  settings  as  conducted   and  proposed
in \cite{amara_graphframex_2024}.  The reason for this  is that  high
$fid_{+}$ scores near  1 are extremely difficult  to achieve compared to
$fid_{-}$ scores near zero. Therefore,  we also consider this  as the preferred
upper  bound in a non-trivial setting.

However,  while the  characterization score  is one  of the  strongest available
explanation evaluation metrics it does  not consider the contextual relationship
between the selected  (C-)CFG and B-AFG. Instead, it only  considers the context
independent explainability of a given graph.  Specifically, the fact that it may
be a related  or important subgraph from a prior  explanation is not considered.
Therefore, we argue  that any ``re-characterization'' of  a highly characterized
subgraph ought  to be especially difficult  to obtain by definition.  That is, a
high characterization  score is indicative of  a selected subgraph that  is both
sufficient and \textit{necessary}.  Therefore, if the subgraph  were observed to
be highly characterized and then  individually re-characterized with a similarly
high value  this would  serve as a  contradiction. As such,  we propose  a level
aware  $\lambda$  explainability  score  that  aims  to  capture  the  level  of
explanation agreement  as the difference  between the (C-)CFG and  B-AFG levels.
Here, a higher value is better, and a negative value is especially unacceptable.

\begin{equation}
  \lambda = charact_{(C-)CFG} - charact_{AFG}
\end{equation}

\subsubsection{Beta Score} Furthermore, we also  recognize that explanations
made at the  (C-)CFG level are not able to  attribute importance at the
instruction assembly  level. This is by definition since nodes that compose
(C-)CFGs  are represented as basic blocks or similar,  i.e., they  consist  of
multiple  instructions that in our case  are aggregated via  a permutation
invariant  function (e.g., sum, mean,  max, etc.). Importantly, this
aggregation is one  way, such  that an explanation  cannot be attributed to  an
individual  instruction post explanation.  To capture  this, we propose a novel
metric, called $\beta$ indicator score,  that outputs a ratio value for a set of
samples indicating the degree of instruction level explainability. $\beta$
returns 0 if  all test samples contain binary values and a positive  value if
any vectors contain non-binary values. If all samples contain non-binary values
$\beta$ is 1.

\begin{equation}
  \beta = \frac{\sum_{i=1}^{n} \mathrm{sgn}\!\left( \sum_{j=1}^{m} \left| x_{ij} (1 - x_{ij} ) \right| \right)}{n}
\end{equation}

We report the $\beta$ value for all  test samples to show if any explanations on
any test sample at the (C-)CFG level  can be used to attribute importance at the
assembly level.  At the B-AFG  level $\beta$ is expected to be 0. We stress the
importance and novelty of this metric since it empirically demonstrates the
level of explanation granularity of any method specific to our setting.

\subsection{Inference   Evaluation}  When   examining  general   trends  over
all experiments at both the  (C-)CFG and B-AFG levels, shown in Table
\ref{tab:inf_results}, we note that most models generally achieve moderate to
good performance (F1 score generally between 85-90\%) in terms of  both positive
and negative  F1 scores (malicious and benign, respectively). However,
consistent under performance  of the  benign class  can be  observed across  all
experiments. This may be due to the ratio of malicious to benign samples,
roughly 1.2 times as many.

\begin{table*}[ht]
\caption{Inference results.}
\centering
\begin{NiceTabular}{ccccccccccc}[hvlines]
\Block{3-1}{Method}          & \Block{3-1}{$r$}                      & \Block{3-1}{Level}                    & \Block{1-8}{Inference}                &                                       &                                       &                                       &                                       &                                       &                                       &                                       \\ 
                             &                                       &                                       & \Block{2-1}{Average Accuracy}             & \Block{2-1}{Accuracy}                 & \Block{1-2}{F1 score}                 &                                       & \Block{1-2}{Recall}                   &                                       & \Block{1-2}{Precision}                &                                       \\ 
                             &                                       &                                       &                                       &                                       & Benign                                & Malicious                             & Benign                                & Malicious                             & Benign                                & Malicious                             \\ 
\Block{2-1}{Baseline}        & \Block[fill=[gray]{0.90}]{2-1}{0}     & \Block[fill=[gray]{0.90}]{1-1}{CFG}   & \Block[fill=[gray]{0.90}]{2-1}{0.894} & \Block[fill=[gray]{0.90}]{1-1}{0.901} & \Block[fill=[gray]{0.90}]{1-1}{0.896} & \Block[fill=[gray]{0.90}]{1-1}{0.907} & \Block[fill=[gray]{0.90}]{1-1}{0.938} & \Block[fill=[gray]{0.90}]{1-1}{0.872} & \Block[fill=[gray]{0.90}]{1-1}{0.857} & \Block[fill=[gray]{0.90}]{1-1}{0.944} \\ 
                             &                                       & \Block[fill=[gray]{0.90}]{1-1}{B-AFG} &                                       & \Block[fill=[gray]{0.90}]{1-1}{0.887} & \Block[fill=[gray]{0.90}]{1-1}{0.882} & \Block[fill=[gray]{0.90}]{1-1}{0.892} & \Block[fill=[gray]{0.90}]{1-1}{0.938} & \Block[fill=[gray]{0.90}]{1-1}{0.846} & \Block[fill=[gray]{0.90}]{1-1}{0.833} & \Block[fill=[gray]{0.90}]{1-1}{0.943} \\ 
\Block{8-1}{Variation Edges} & \Block{2-1}{0.25}                     & C-CFG                                 & \Block{2-1}{\textbf{0.898}}           & 0.894                                 & 0.884                                 & 0.903                                 & 0.891                                 & 0.897                                 & 0.877                                 & 0.909                                 \\ 
                             &                                       & B-AFG                                 &                                       & 0.901                                 & 0.894                                 & 0.908                                 & 0.922                                 & 0.885                                 & 0.868                                 & 0.932                                 \\ 
                             & \Block[fill=[gray]{0.90}]{2-1}{0.5}   & \Block[fill=[gray]{0.90}]{1-1}{C-CFG} & \Block[fill=[gray]{0.90}]{2-1}{0.891} & \Block[fill=[gray]{0.90}]{1-1}{0.901} & \Block[fill=[gray]{0.90}]{1-1}{0.883} & \Block[fill=[gray]{0.90}]{1-1}{0.915} & \Block[fill=[gray]{0.90}]{1-1}{0.828} & \Block[fill=[gray]{0.90}]{1-1}{0.962} & \Block[fill=[gray]{0.90}]{1-1}{0.946} & \Block[fill=[gray]{0.90}]{1-1}{0.872} \\ 
                             &                                       & \Block[fill=[gray]{0.90}]{1-1}{B-AFG} &                                       & \Block[fill=[gray]{0.90}]{1-1}{0.880} & \Block[fill=[gray]{0.90}]{1-1}{0.864} & \Block[fill=[gray]{0.90}]{1-1}{0.893} & \Block[fill=[gray]{0.90}]{1-1}{0.844} & \Block[fill=[gray]{0.90}]{1-1}{0.910} & \Block[fill=[gray]{0.90}]{1-1}{0.885} & \Block[fill=[gray]{0.90}]{1-1}{0.877} \\ 
                             & \Block{2-1}{0.75}                     & C-CFG                                 & \Block{2-1}{0.894}                    & 0.923                                 & 0.917                                 & 0.927                                 & 0.953                                 & 0.897                                 & 0.884                                 & 0.959                                 \\ 
                             &                                       & B-AFG                                 &                                       & 0.866                                 & 0.861                                 & 0.871                                 & 0.922                                 & 0.821                                 & 0.808                                 & 0.928                                 \\ 
                             & \Block[fill=[gray]{0.90}]{2-1}{0.999} & \Block[fill=[gray]{0.90}]{1-1}{C-CFG} & \Block[fill=[gray]{0.90}]{2-1}{0.845} & \Block[fill=[gray]{0.90}]{1-1}{0.810} & \Block[fill=[gray]{0.90}]{1-1}{0.803} & \Block[fill=[gray]{0.90}]{1-1}{0.816} & \Block[fill=[gray]{0.90}]{1-1}{0.859} & \Block[fill=[gray]{0.90}]{1-1}{0.769} & \Block[fill=[gray]{0.90}]{1-1}{0.753} & \Block[fill=[gray]{0.90}]{1-1}{0.870} \\ 
                             &                                       & \Block[fill=[gray]{0.90}]{1-1}{B-AFG} &                                       & \Block[fill=[gray]{0.90}]{1-1}{0.880} & \Block[fill=[gray]{0.90}]{1-1}{0.866} & \Block[fill=[gray]{0.90}]{1-1}{0.892} & \Block[fill=[gray]{0.90}]{1-1}{0.859} & \Block[fill=[gray]{0.90}]{1-1}{0.897} & \Block[fill=[gray]{0.90}]{1-1}{0.873} & \Block[fill=[gray]{0.90}]{1-1}{0.886} \\ 
\Block{8-1}{Kron}            & \Block{2-1}{0.25}                     & C-CFG                                 & \Block{2-1}{0.877}                    & 0.880                                 & 0.870                                 & 0.889                                 & 0.891                                 & 0.872                                 & 0.851                                 & 0.907                                 \\ 
                             &                                       & B-AFG                                 &                                       & 0.873                                 & 0.862                                 & 0.883                                 & 0.875                                 & 0.872                                 & 0.848                                 & 0.895                                 \\ 
                             & \Block[fill=[gray]{0.90}]{2-1}{0.5}   & \Block[fill=[gray]{0.90}]{1-1}{C-CFG} & \Block[fill=[gray]{0.90}]{2-1}{0.866} & \Block[fill=[gray]{0.90}]{1-1}{0.880} & \Block[fill=[gray]{0.90}]{1-1}{0.876} & \Block[fill=[gray]{0.90}]{1-1}{0.884} & \Block[fill=[gray]{0.90}]{1-1}{0.938} & \Block[fill=[gray]{0.90}]{1-1}{0.833} & \Block[fill=[gray]{0.90}]{1-1}{0.822} & \Block[fill=[gray]{0.90}]{1-1}{0.942} \\ 
                             &                                       & \Block[fill=[gray]{0.90}]{1-1}{B-AFG} &                                       & \Block[fill=[gray]{0.90}]{1-1}{0.852} & \Block[fill=[gray]{0.90}]{1-1}{0.847} & \Block[fill=[gray]{0.90}]{1-1}{0.857} & \Block[fill=[gray]{0.90}]{1-1}{0.906} & \Block[fill=[gray]{0.90}]{1-1}{0.808} & \Block[fill=[gray]{0.90}]{1-1}{0.795} & \Block[fill=[gray]{0.90}]{1-1}{0.913} \\ 
                             & \Block{2-1}{0.75}                     & C-CFG                                 & \Block{2-1}{\textbf{0.898}}           & 0.894                                 & 0.882                                 & 0.904                                 & 0.875                                 & 0.910                                 & 0.889                                 & 0.899                                 \\ 
                             &                                       & B-AFG                                 &                                       & 0.901                                 & 0.892                                 & 0.909                                 & 0.906                                 & 0.897                                 & 0.879                                 & 0.921                                 \\ 
                             & \Block[fill=[gray]{0.90}]{2-1}{0.999} & \Block[fill=[gray]{0.90}]{1-1}{C-CFG} & \Block[fill=[gray]{0.90}]{2-1}{0.863} & \Block[fill=[gray]{0.90}]{1-1}{0.831} & \Block[fill=[gray]{0.90}]{1-1}{0.818} & \Block[fill=[gray]{0.90}]{1-1}{0.842} & \Block[fill=[gray]{0.90}]{1-1}{0.844} & \Block[fill=[gray]{0.90}]{1-1}{0.821} & \Block[fill=[gray]{0.90}]{1-1}{0.794} & \Block[fill=[gray]{0.90}]{1-1}{0.865} \\ 
                             &                                       & \Block[fill=[gray]{0.90}]{1-1}{B-AFG} &                                       & \Block[fill=[gray]{0.90}]{1-1}{0.894} & \Block[fill=[gray]{0.90}]{1-1}{0.878} & \Block[fill=[gray]{0.90}]{1-1}{0.907} & \Block[fill=[gray]{0.90}]{1-1}{0.844} & \Block[fill=[gray]{0.90}]{1-1}{0.936} & \Block[fill=[gray]{0.90}]{1-1}{0.915} & \Block[fill=[gray]{0.90}]{1-1}{0.880} \\ 
\end{NiceTabular}
\label{tab:inf_results}
\end{table*}

\begin{figure}[h]
     \centering
     \includegraphics[width=0.5\textwidth]{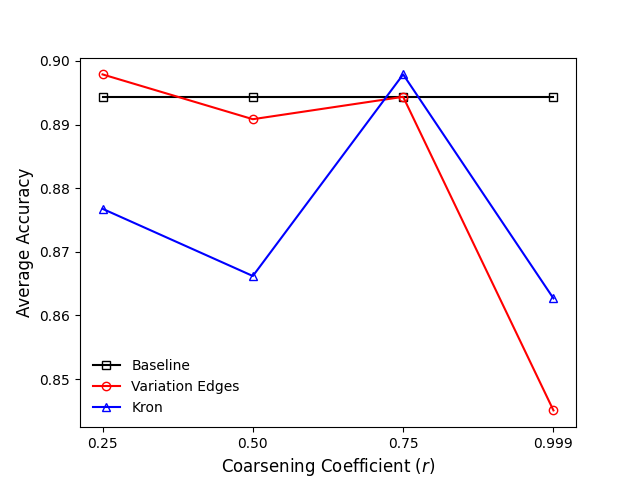}
     \caption{Meta-Coarsening average accuracy.}
    \label{fig:results_plots}
\end{figure}

When   assessing  inference   performance   of   our  Meta-Coarsening   approach
holistically,  shown  in Fig. \ref{fig:results_plots}, we consider the average
accuracy of  both C-CFG  and B-AFG levels together. Here, we observe that both
coarsening methods  decline from low to mild ($r$ of 0.25 to 0.50) coarsening,
increase from mild to moderate coarsening ($r$ of 0.50 to 0.75), and significant
drop from moderate to maximal coarsening ($r$ of 0.75 to 0.999). Furthermore, we
observe that Variation Edges is consistently closer to the baseline than Kron.
The best performance of 89.8\%, \textbf{+0.4} over the baseline, is achieved by
both Kron and Variation edges with $r$ values of 0.25 and 0.75, respectively.
However, we recognize that Kron is better since it achieves the best performance
at a higher coarsening level making it more computationally efficient for GNN
training and explanation. 25\% CFG size compared to 75\% CFG size in the case of
Kron. Overall, this aligns with our our intuition that coarsening may act to
regularize samples and make it easier for a GNN model to generalize. However,
with respect to \textbf{RQ1}, certain coarsening methods and coefficients can be
more beneficial for inference performance than others and improve over the
baseline.

When assessing the performance of each level individually, shown in Table
\ref{tab:inf_results}, the highest  C-CFG level  accuracy is achieved by
Variation Edges with $r$ of 0.75 with  92.3\% accuracy, \textbf{+2.2\%} over the
baseline. Furthermore, the best B-AFG level inference performance is achieved by
both Kron and Variation Edges at $r$ values of 0.25 and 0.75, respectively, that
achieve 90.1\% accuracy, \textbf{+1.4\%} over the baseline. This slight drop in
inference performance  between the C-CFG and B-AFG levels matches our intuition
since B-AFG graphs are more variable and contain  less information compared to
the (C-)CFG level. With respect to individual coarsening  trends generally, we
observe an increase in performance at the C-CFG level  for both methods as
coarsening is increased, and, then,  a significant drop once maximal coarsening
is reached. This suggests that a coarsening threshold exists where samples may
be overly regularized making it hard for a GNN model to distinguish them.
However,  at the B-AFG level  the overall trend of  coarsening, while far less
clear  is noticeably  more consistent. Ranging between  85.2-90.1\% as compared
to 81.0-92.3\% at the C-CFG level. Furthermore, we also observe that maximal
coarsening also decreases performance similarly for both methods. 

When evaluating classwise F1 scores at the C-CFG and B-AFG levels,  shown in
Table \ref{tab:inf_results}, we observe similar trends as in panels \textit{b}
and \textit{c}, respectively. As mentioned prior, we also observe under
consistent under performance in the benign class (approximately 2-3\%) across
varying $r$ values and the baseline. Overall, these trends suggest that class
specific inference is not extremely impacted by coarsening, thus addressing the
inference aspect of \textbf{RQ3}.

Lastly, inference evaluation  for the proposed Meta-Coarsening approach  should
be viewed and considered  with respect to  its intended purpose; to  provide
comprehensive supporting evidence of model performance  towards our
explainability results. The overall goal  is not for  our Meta-Coarsening
approach  to be used  strictly for inference but to support malware detection
and assembly level explainability.

\begin{table*}[ht]
\caption{Explainability results. Combined malicious and benign classes denoted with \textit{All}.}
\centering
\begin{NiceTabular}{cccccccccc}[hvlines]
\Block{3-1}{Method}          & \Block{3-1}{$r$}                      & \Block{3-1}{Level}                    & \Block{1-7}{Explainability}           &                                       &                                       &                                       &                                          &                                       & \\
                             &                                       &                                       & \Block{1-4}{Average Characterization}              &                                       &                                       &                                       & \Block{1-2}{Average Fidelity\textsubscript{\textit{All}}} &                                       & \Block{2-1}{$\beta$} \\
                             &                                       &                                       & $\lambda_\text{\textit{All}}$                       & \textit{All}                                   & Benign                                & Malicious                             & Minus                                    & Plus                                  & \\
\Block{2-1}{Baseline}        & \Block[fill=[gray]{0.90}]{2-1}{0}     & \Block[fill=[gray]{0.90}]{1-1}{CFG}   & \Block[fill=[gray]{0.90}]{2-1}{0.637} & \Block[fill=[gray]{0.90}]{1-1}{0.724} & \Block[fill=[gray]{0.90}]{1-1}{0.805} & \Block[fill=[gray]{0.90}]{1-1}{0.587} & \Block[fill=[gray]{0.90}]{1-1}{0.106} & \Block[fill=[gray]{0.90}]{1-1}{0.621} & \Block[fill=[gray]{0.90}]{1-1}{1} \\
                             &                                       & \Block[fill=[gray]{0.90}]{1-1}{B-AFG} &                                       & \Block[fill=[gray]{0.90}]{1-1}{0.087} & \Block[fill=[gray]{0.90}]{1-1}{0.029} & \Block[fill=[gray]{0.90}]{1-1}{0.133} & \Block[fill=[gray]{0.90}]{1-1}{0.031} & \Block[fill=[gray]{0.90}]{1-1}{0.046} & \Block[fill=[gray]{0.90}]{1-1}{\textbf{0}} \\
\Block{8-1}{Variation Edges} & \Block{2-1}{0.25}                     & C-CFG                                 & \Block{2-1}{0.590}                    & 0.692                                 & 0.751                                 & 0.515                                 & 0.153                                 & 0.594                                 & 1                                 \\
                             &                                       & B-AFG                                 &                                       & 0.102                                 & 0.040                                 & 0.151                                 & 0.047                                 & 0.054                                 & \textbf{0}                                 \\
                             & \Block[fill=[gray]{0.90}]{2-1}{0.5}   & \Block[fill=[gray]{0.90}]{1-1}{C-CFG} & \Block[fill=[gray]{0.90}]{2-1}{0.532} & \Block[fill=[gray]{0.90}]{1-1}{0.684} & \Block[fill=[gray]{0.90}]{1-1}{0.758} & \Block[fill=[gray]{0.90}]{1-1}{0.560} & \Block[fill=[gray]{0.90}]{1-1}{0.106} & \Block[fill=[gray]{0.90}]{1-1}{0.567} & \Block[fill=[gray]{0.90}]{1-1}{1} \\
                             &                                       & \Block[fill=[gray]{0.90}]{1-1}{B-AFG} &                                       & \Block[fill=[gray]{0.90}]{1-1}{0.152} & \Block[fill=[gray]{0.90}]{1-1}{0.106} & \Block[fill=[gray]{0.90}]{1-1}{0.185} & \Block[fill=[gray]{0.90}]{1-1}{0.096} & \Block[fill=[gray]{0.90}]{1-1}{0.084} & \Block[fill=[gray]{0.90}]{1-1}{\textbf{0}} \\
                             & \Block{2-1}{0.75}                     & C-CFG                                 & \Block{2-1}{0.517}                    & 0.675                                 & 0.761                                 & 0.518                                 & 0.154                                 & 0.576                                 & 1                                 \\
                             &                                       & B-AFG                                 &                                       & 0.158                                 & 0.167                                 & 0.149                                 & 0.111                                 & 0.091                                 & \textbf{0}                                 \\
                             & \Block[fill=[gray]{0.90}]{2-1}{0.999} & \Block[fill=[gray]{0.90}]{1-1}{C-CFG} & \Block[fill=[gray]{0.90}]{2-1}{0.190} & \Block[fill=[gray]{0.90}]{1-1}{0.501} & \Block[fill=[gray]{0.90}]{1-1}{0.284} & \Block[fill=[gray]{0.90}]{1-1}{0.000} & \Block[fill=[gray]{0.90}]{1-1}{0.455} & \Block[fill=[gray]{0.90}]{1-1}{0.471} & \Block[fill=[gray]{0.90}]{1-1}{1} \\
                             &                                       & \Block[fill=[gray]{0.90}]{1-1}{B-AFG} &                                       & \Block[fill=[gray]{0.90}]{1-1}{0.312} & \Block[fill=[gray]{0.90}]{1-1}{0.349} & \Block[fill=[gray]{0.90}]{1-1}{0.137} & \Block[fill=[gray]{0.90}]{1-1}{0.301} & \Block[fill=[gray]{0.90}]{1-1}{0.216} & \Block[fill=[gray]{0.90}]{1-1}{\textbf{0}} \\
\Block{8-1}{Kron}            & \Block{2-1}{0.25}                     & C-CFG                                 & \Block{2-1}{\textbf{0.713}}                    & 0.770                                 & 0.790                                 & 0.722                                 & 0.083                                 & 0.675                                 & 1                                 \\
                             &                                       & B-AFG                                 &                                       & 0.057                                 & 0.049                                 & 0.063                                 & 0.042                                 & 0.030                                 & \textbf{0}                                 \\
                             & \Block[fill=[gray]{0.90}]{2-1}{0.5}   & \Block[fill=[gray]{0.90}]{1-1}{C-CFG} & \Block[fill=[gray]{0.90}]{2-1}{0.599} & \Block[fill=[gray]{0.90}]{1-1}{0.707} & \Block[fill=[gray]{0.90}]{1-1}{0.834} & \Block[fill=[gray]{0.90}]{1-1}{0.565} & \Block[fill=[gray]{0.90}]{1-1}{0.066} & \Block[fill=[gray]{0.90}]{1-1}{0.586} & \Block[fill=[gray]{0.90}]{1-1}{1} \\
                             &                                       & \Block[fill=[gray]{0.90}]{1-1}{B-AFG} &                                       & \Block[fill=[gray]{0.90}]{1-1}{0.107} & \Block[fill=[gray]{0.90}]{1-1}{0.080} & \Block[fill=[gray]{0.90}]{1-1}{0.128} & \Block[fill=[gray]{0.90}]{1-1}{0.122} & \Block[fill=[gray]{0.90}]{1-1}{0.060} & \Block[fill=[gray]{0.90}]{1-1}{\textbf{0}} \\
                             & \Block{2-1}{0.75}                     & C-CFG                                 & \Block{2-1}{0.455}                    & 0.634                                 & 0.730                                 & 0.480                                 & 0.123                                 & 0.509                                 & 1                                 \\
                             &                                       & B-AFG                                 &                                       & 0.179                                 & 0.195                                 & 0.163                                 & 0.098                                 & 0.104                                 & 
                             \textbf{0}                                 \\
                             & \Block[fill=[gray]{0.90}]{2-1}{0.999} & \Block[fill=[gray]{0.90}]{1-1}{C-CFG} & \Block[fill=[gray]{0.90}]{2-1}{0.151} & \Block[fill=[gray]{0.90}]{1-1}{0.400} & \Block[fill=[gray]{0.90}]{1-1}{0.229} & \Block[fill=[gray]{0.90}]{1-1}{0.007} & \Block[fill=[gray]{0.90}]{1-1}{0.505} & \Block[fill=[gray]{0.90}]{1-1}{0.361} & \Block[fill=[gray]{0.90}]{1-1}{1} \\
                             &                                       & \Block[fill=[gray]{0.90}]{1-1}{B-AFG} &                                       & \Block[fill=[gray]{0.90}]{1-1}{0.249} & \Block[fill=[gray]{0.90}]{1-1}{0.327} & \Block[fill=[gray]{0.90}]{1-1}{0.164} & \Block[fill=[gray]{0.90}]{1-1}{0.167} & \Block[fill=[gray]{0.90}]{1-1}{0.155} & \Block[fill=[gray]{0.90}]{1-1}{\textbf{0}} \\
\end{NiceTabular}
\label{tab:exp_results}
\end{table*}

\subsection{Explainability Evaluation}  While  typical inference performance
metrics are  computed  using the  entire  test set, explainability  metrics
require evaluation  using correctly  predicted test  samples only.  This is  due
to  the focus of  understanding the  phenomenon related to  an aspect of class
decidability rather  than evaluating  the decidability of  the GNN model itself
\cite{amara_graphframex_2024}.  Therefore,  we only consider correctly predicted
test samples when computing explainability metrics.

\begin{figure}[h]
     \centering
     \includegraphics[width=0.5\textwidth]{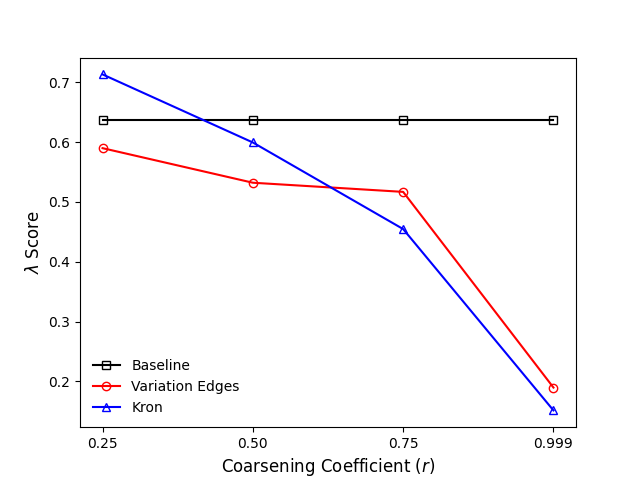}
     \caption{$\lambda$ scores.}
    \label{fig:exp_lam_and_charct}
\end{figure}

\begin{figure}[h]
     \centering
     \includegraphics[width=0.5\textwidth]{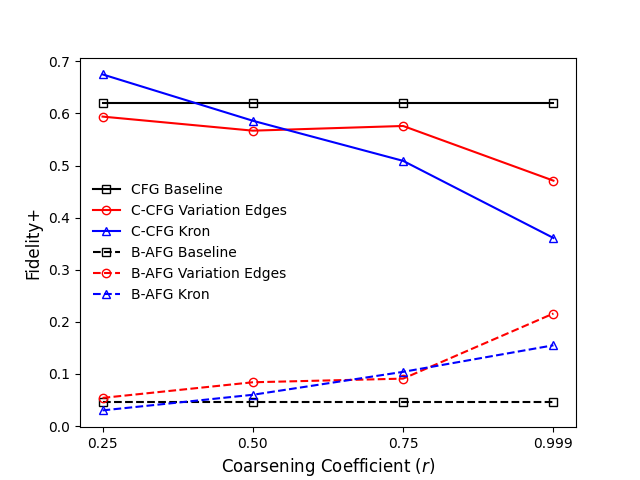}
     \caption{$fid_{+}$ results at (C-)CFG and B-AFG levels.}
    \label{fig:exp_fidelity_plus}
\end{figure}

In terms of  explainability, shown in Table \ref{tab:exp_results}  and
summarized in Fig.  \ref{fig:exp_lam_and_charct}, we observe  that  Kron with a
coarsening  $r$  value  of 0.25  achieves  the highest $\lambda$  score of
\textbf{0.713},  indicating  strong  explainability agreement  between  the
C-CFG  and  B-AFG.   However,  fully interpreting  this score  requires  further
assessing the fidelity components ($fid_{+}$  and $fid_{-}$) of  the respective
characterization  scores. Most importantly,  we observe that  $fid_{+}$ values,
shown  in  Fig. \ref{fig:exp_fidelity_plus}, are the main result of this
characterization difference from 0.675 at the C-CFG level  to 0.030  at the
B-AFG  level.  These  $fid_{+}$ scores strongly suggest that our  selected C-CFG
subgraphs are \textit{necessary}. Meaning that their removal causes
misclassification. Furthermore, low values at the B-AFG level indicate that  it
is extremely difficult  to remove a subgraph  to force misclassification, albeit
at a different but  highly related level. A low score at the  B-AFG level  also
serves to  confirm the prior $fid_{+}$ score  used to derive the B-AFG.

\begin{figure}[h]
     \centering
     \includegraphics[width=0.5\textwidth]{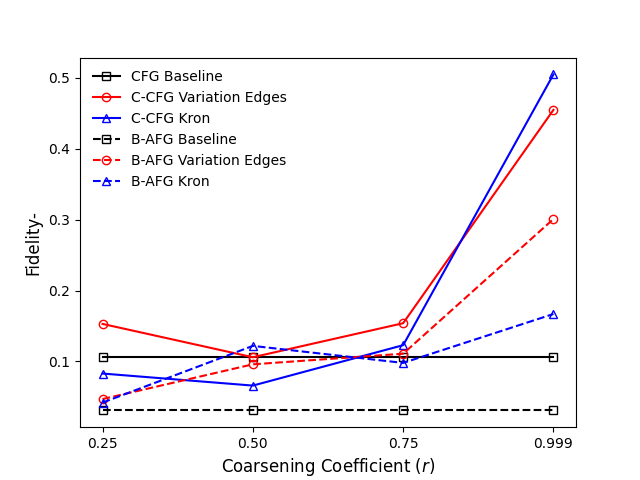}
     \caption{$fid_{-}$ results at (C-)CFG and B-AFG levels.}
    \label{fig:exp_fidelity_minus}
\end{figure}

Furthermore, we  also observe  a very  small increase  in $fid_{-}$  values from
0.083  at  the  C-CFG level  to  0.042  at  the  B-AFG level,  shown  in  Fig.
\ref{fig:exp_fidelity_minus}. These $fid_{-}$ scores  strongly indicate that our
selected subgraphs, at both levels, are  highly \textit{sufficient}.  Meaning
that  they provide  the same correct predictions.

\begin{figure*}[h!]
    \centering
    \subfloat[Coarsened CFG.]{
        \includegraphics[width=0.30\textwidth]{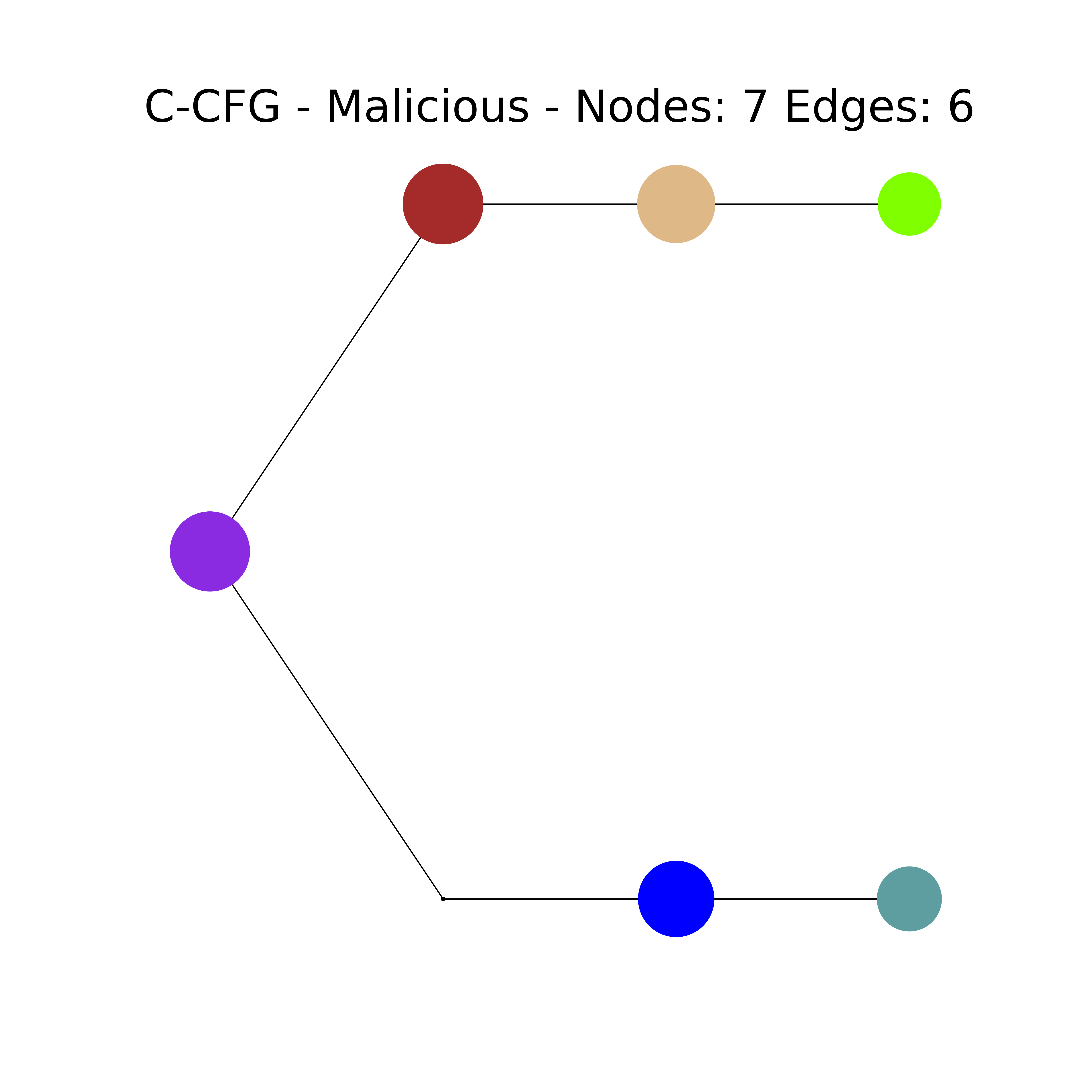}
        \label{fig:sub_a}
    } \hfil
    \subfloat[CFG with supernode color backtracking from panel \textit{a}.]{
        \includegraphics[width=0.30\textwidth]{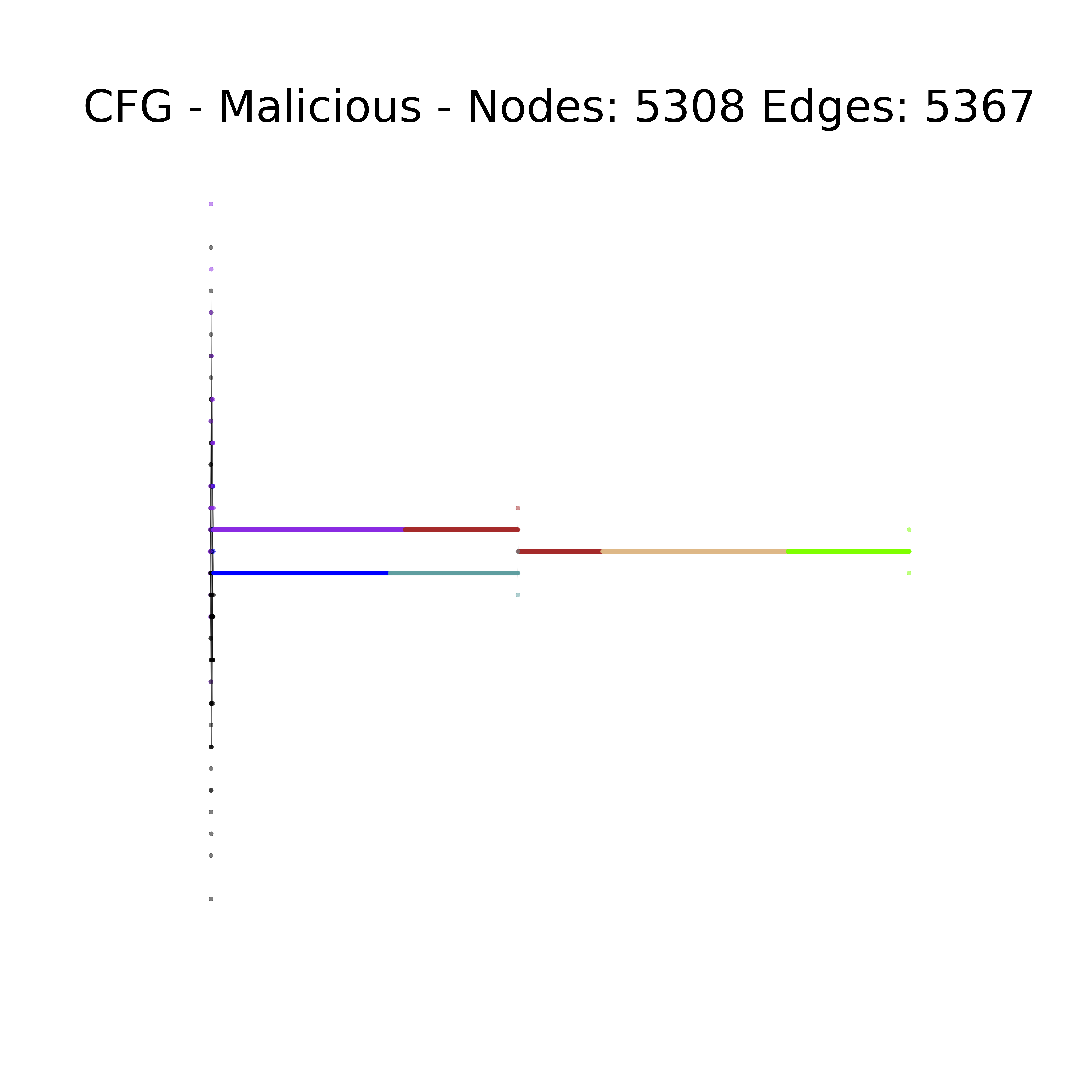}
        \label{fig:sub_b}
    } \hfil
    \subfloat[AFG with supernode color backtracking from panel \textit{a} via panel \textit{b}.]{
        \includegraphics[width=0.30\textwidth]{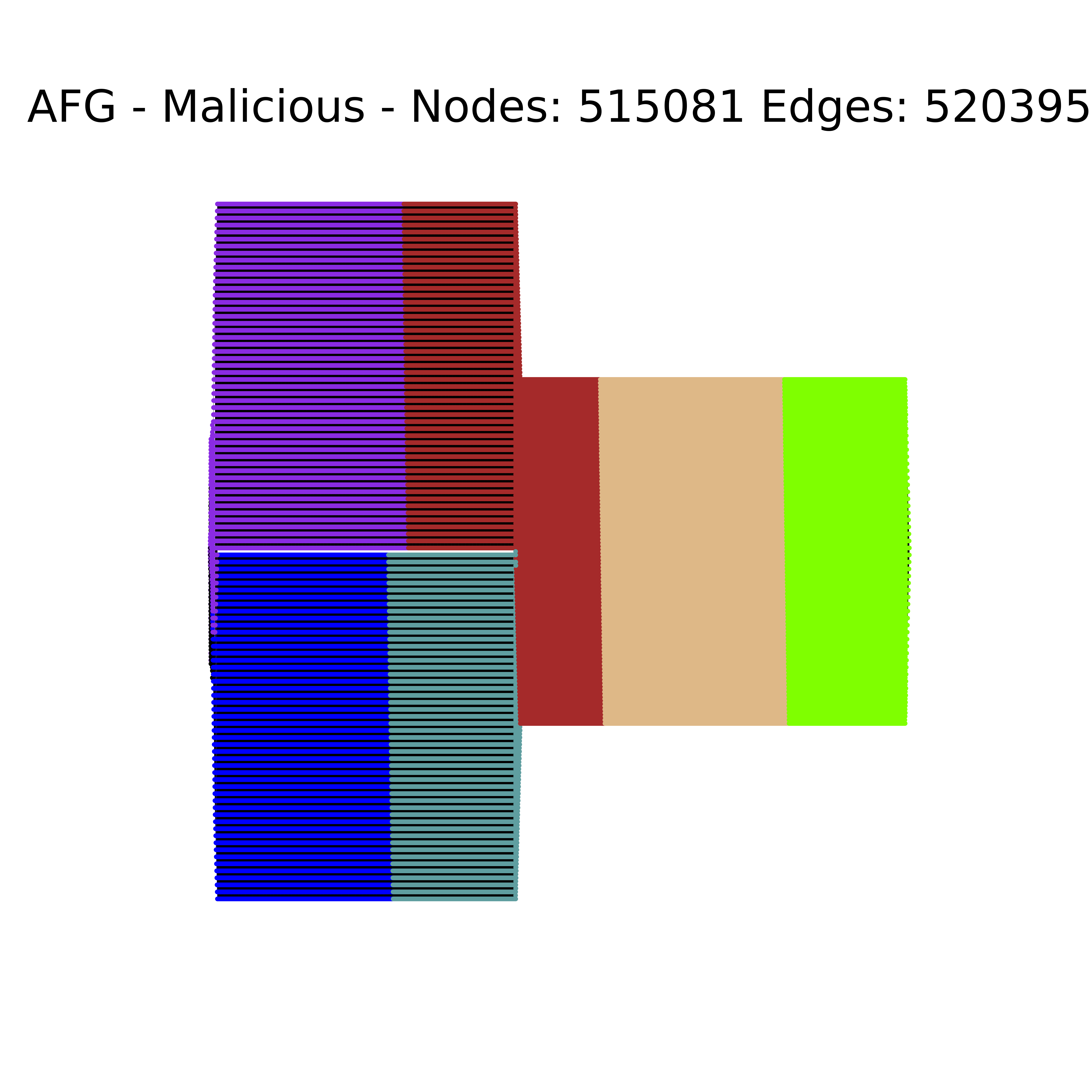}
        \label{fig:sub_c}
    } \\
    \subfloat[Selected Coarsened CFG induced via IG-GEC and $\epsilon$ selection on panel \textit{g}.]{
        \includegraphics[width=0.30\textwidth]{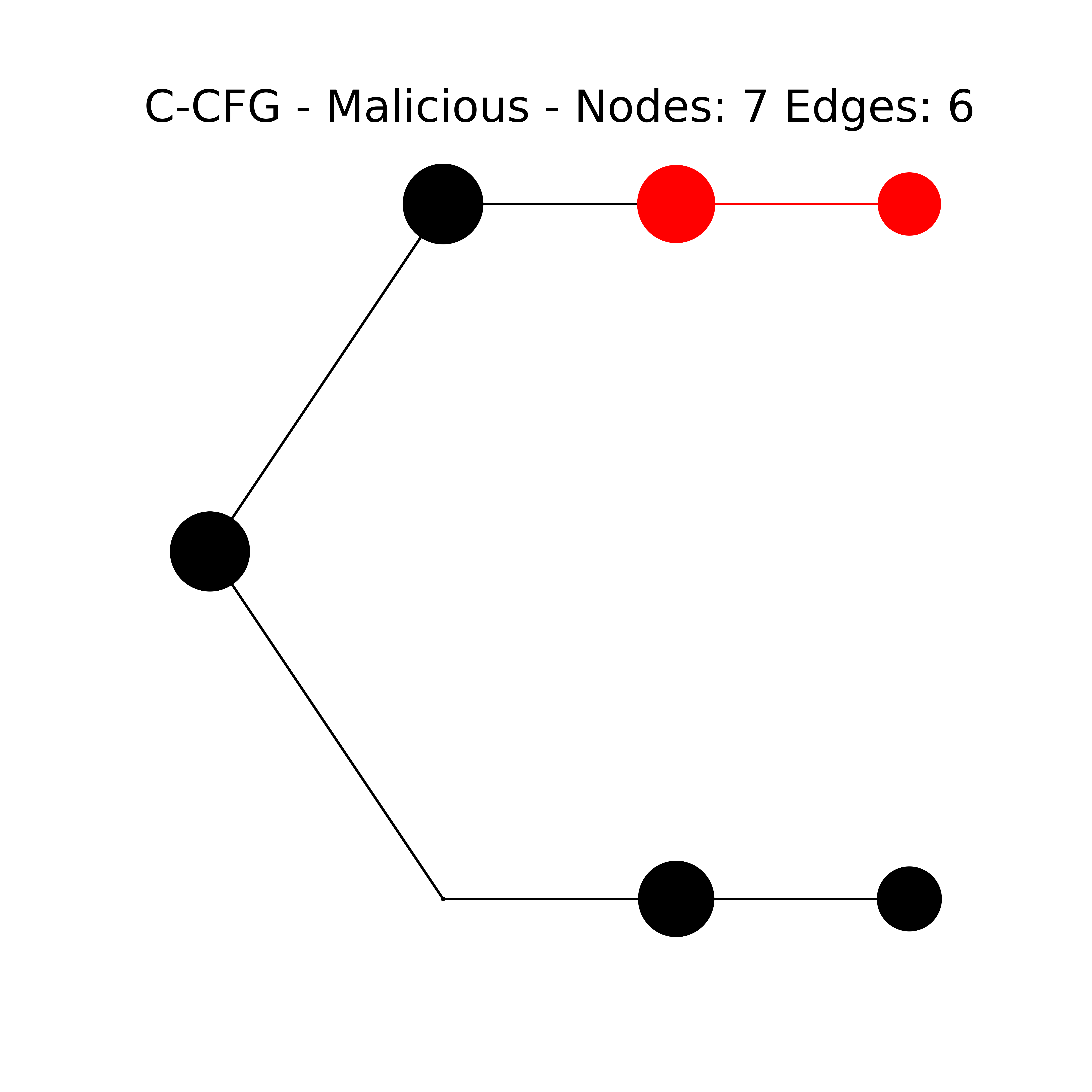}
        \label{fig:sub_d}
    } \hfil
    \subfloat[Selected CFG backtracked from panel \textit{d}.]{
        \includegraphics[width=0.30\textwidth]{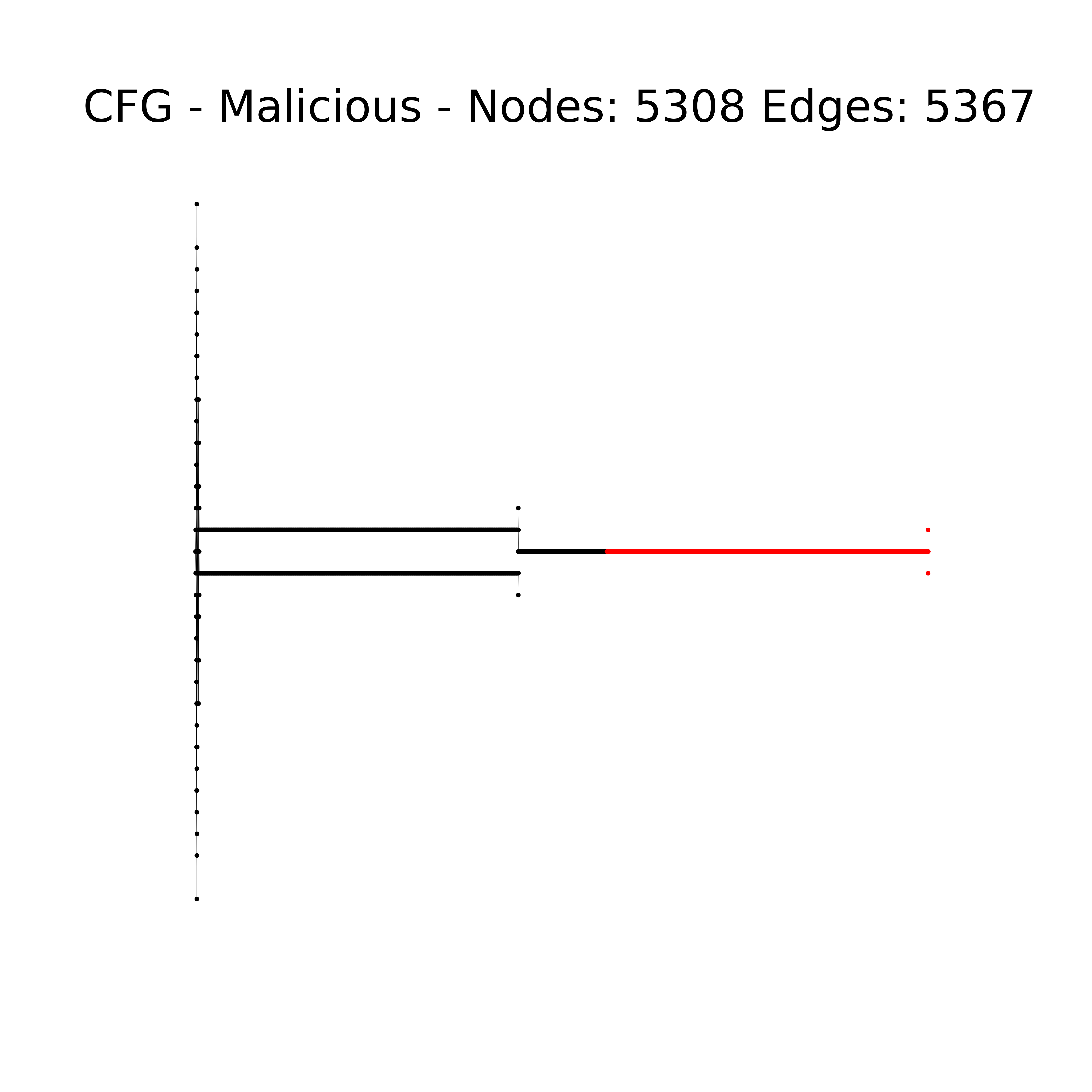}
        \label{fig:sub_e}
    } \hfil
    \subfloat[Selected AFG backtracked from panel \textit{d} via panel \textit{e}.]{
        \includegraphics[width=0.30\textwidth]{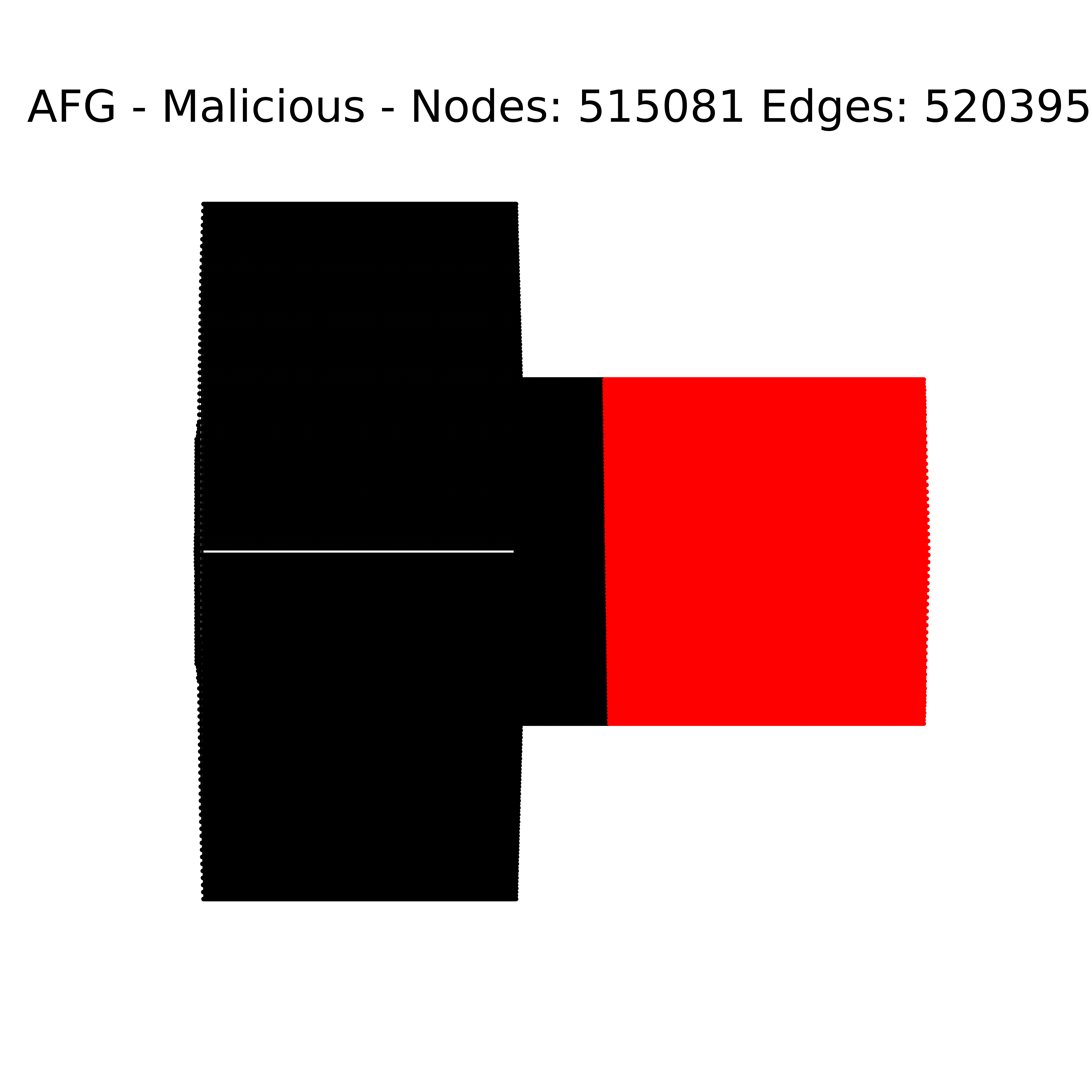}
        \label{fig:sub_f}
    } \\
    \subfloat[Explained (C-)CFG.]{
        \includegraphics[width=0.30\textwidth]{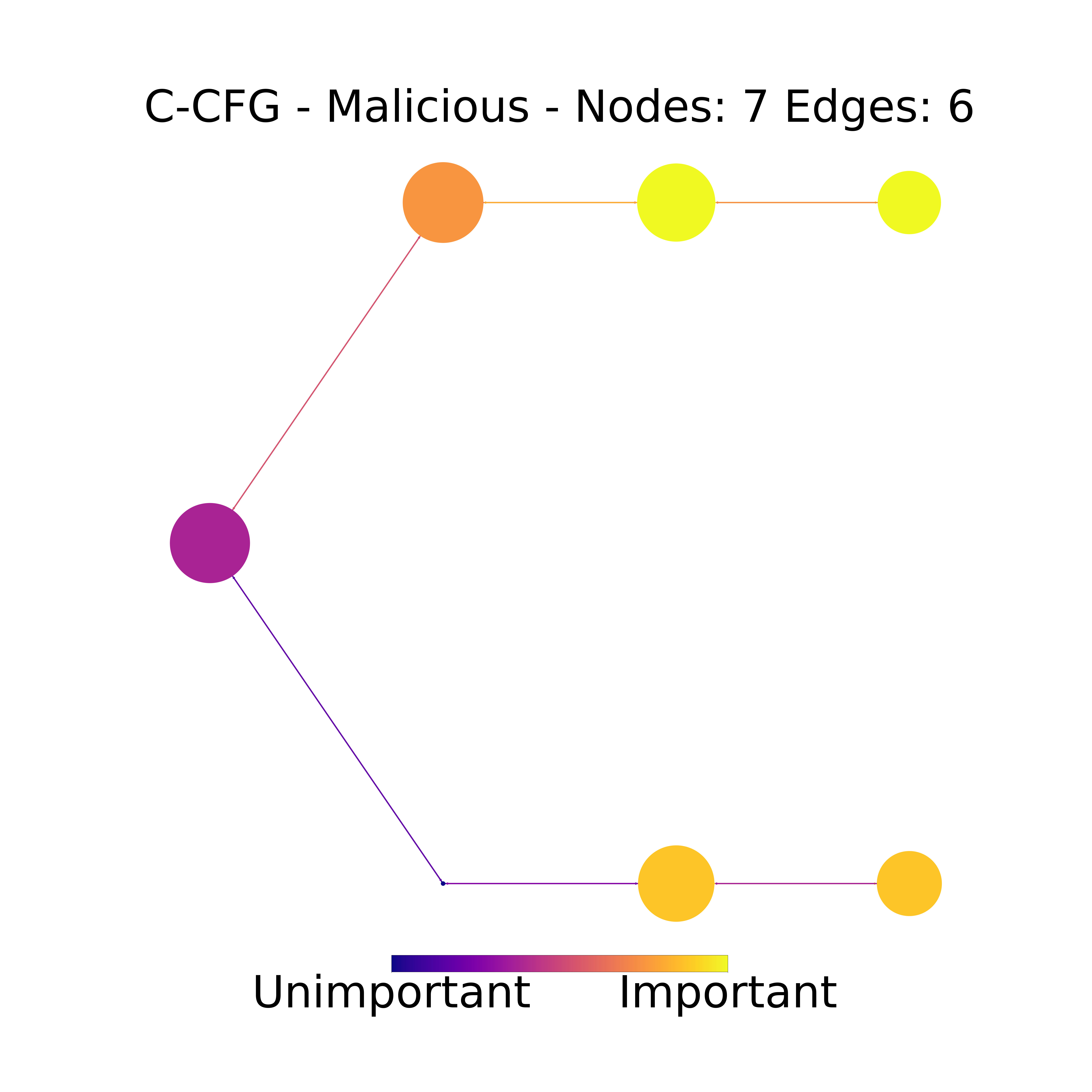}
        \label{fig:sub_g}
    } \hfil
    \subfloat[Explained AFG backtracked from panel \textit{f}.]{
        \includegraphics[width=0.30\textwidth]{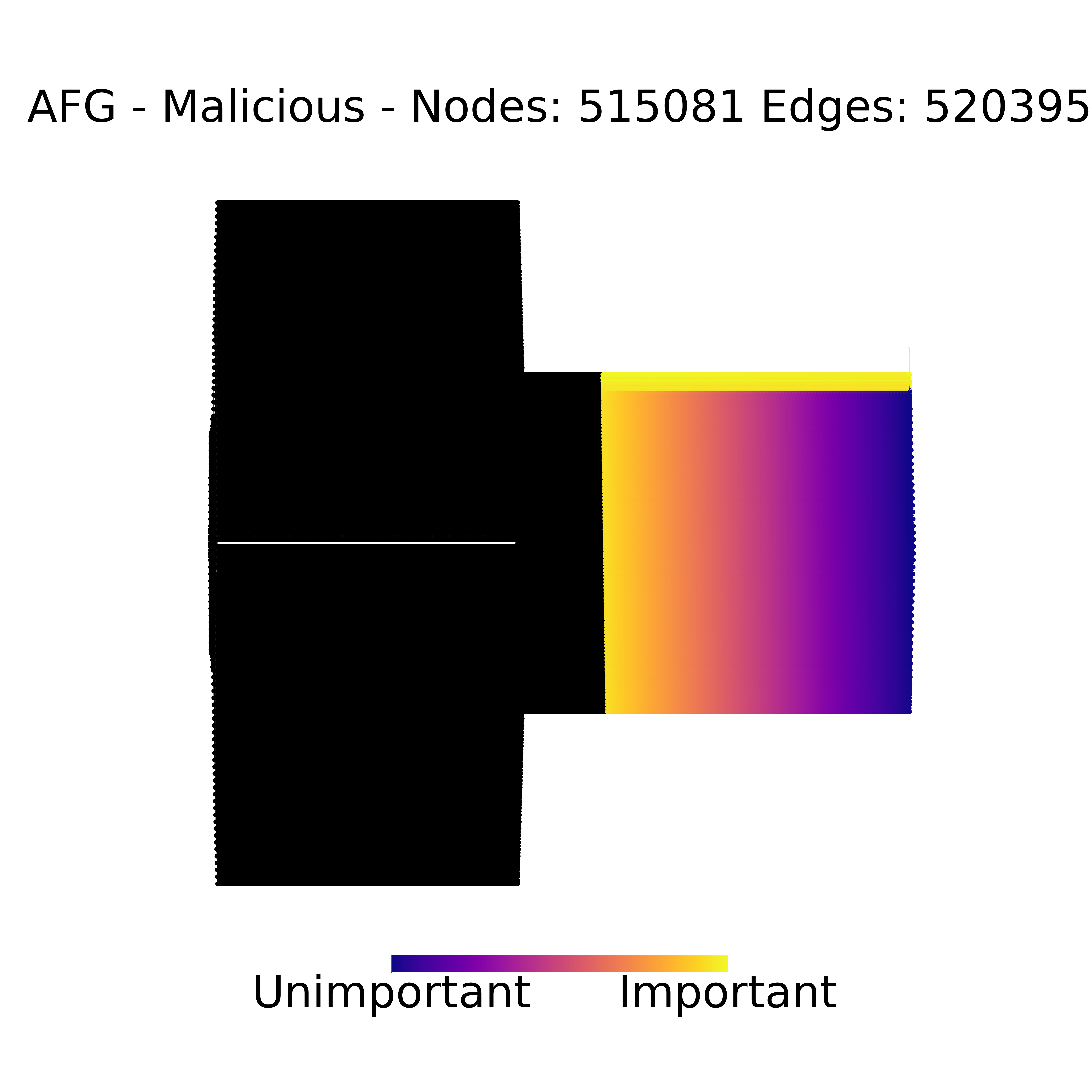}
        \label{fig:sub_h}
    }
    \caption{Plots of coarsened, selected, and explained malicious graph at the (C-)CFG and (B-)AFG levels with Kron coarsening ($r=0.999$), IG explainer, $\tau=\text{TES}$, and $\epsilon=10\%$. Node size in all panels are representative of the total number of instructions that belong to it. Hot and cool color gradients (yellow to purple) in panels \textit{g} and \textit{h} represent explained importance from high to low, respectively.}
    \label{fig:meta_coarsening_vis_max}
\end{figure*}

Overall,  these  findings indicate  that  only  $fid_{+}$ is  strongly  impacted
by  our  Meta-Coarsening  method.  Further  evidence  of our Meta-Coarsening
approach to provide  strong characterization capability for malicious and benign
phenomenon, thus addressing \textbf{RQ2}.

Furthermore,  as coarsening  is increased we generally  observe  a decrease  in
$\lambda$  score. This aligns with  our  intuition since   coarsening  focus on
contracting nodes with respect to graph structure rather  than importance. This
leads   to reduced differentiability since important and unimportant nodes
become non-separable when contracted into supernodes. Unlike inference, this
naturally harms explainability since  separability is an important aspect of
deriving post-hoc explainability. Additionally,  we also  observe  significant
differences  in classwise $\lambda$  and characterization scores, shown  in
Table \ref{tab:exp_results}. Specifically, a baseline \textbf{31.2} point
$\lambda$ difference exists between benign and malicious classes. This  suggests
that the malicious class is generally much  less explainable than the  benign
class overall in  our setting. Furthermore, when coarsening is applied, we
observe that explainability for both classes degrades with respect to  the
baseline case while maintaining the prior difference. The exception being Kron
with $r$ of 0.25 that interestingly boosts malicious explainability over the
baseline. Furthermore, when coarsening  is maximal explainability  is reduced
for both classes below zero indicating extremely poor explainability. These
observations help address \textbf{RQ3} in terms of explainability.

Most importantly, we  also observe  that the  $\beta$ indicator  is maximal for
all  (C-)CFGs and minimal for all B-AFGs, shown in Table \ref{tab:exp_results}.
This means that there is no ambiguity with respect to the  explanations at the
B-AFG level since they are made at the assembly instruction  level compared to
(C-)CFG, where there  is always ambiguity at the basic block level. \textbf{This
finding is extremely important as  it empirically highlights a major limitation
of CFG or similar (e.g., FCG)  with  respect to explanation  granularity that
our proposed approach addresses}.

Lastly,   we   visualize   an    end-to-end   test   sample   using   our
Meta-Coarsening approach, shown in Fig.  \ref{fig:meta_coarsening_vis_max},
where we  consider  Kron  with  $r=0.999$ to assist with visualization. In all
panels, the  size of nodes indicate the number of instructions present. In panel
\textit{a}, the C-CFG is visualized with nodes of differing colors representing
the supernodes backtracked onto the  CFG in panel \textit{b} and B-AFG in panel
\textit{c}. This shows that the coarsened supernode mapping is consistent across
the various levels.  Next, in panel \textit{g}, we visualize the  C-CFG
explainability.  Here, yellow  nodes are  more important  compared to orange and
purple nodes, respectively. In  panel \textit{d}, we visualize  the top selected
nodes  using the explanation  from panel \textit{g}. We  also visualize the
backtracking from the CFG level onto the B-AFG level in panels  \textit{e}  and
\textit{f}, respectively.  This  also demonstrates  that selection  mapping is
consistent across  the various  levels. Lastly,  in panel \textit{h}, we
visualize the explainability at  the B-AFG level. Here, we observe different
regions, some with smooth gradients indicating their importance.

\section{Discussion and Limitations}
\label{sec:discussion_and_limitations}

With respect to \textbf{RQ1}, the application of coarsening is beneficial for
inference performance using specific coarsening methods and coefficients,
i.e. Variation Edges and Kron with $r$ values of 0.25 and 0.75, respectively.

Regarding \textbf{RQ2}, the impact of  our proposed Meta-Coarsening approach can
improve explainability at  low to moderate coarsening levels. Specifically, we
recognize that our $\lambda$  score,  that considers  the dependency between
(C-)CFG and B-AFG levels, shows a large increase  over the  baseline  case  for
low coarsening  with Kron specifically. However,  high   to  maximal coarsening
tends  to   diminish explainability. Furthermore,  our  $\beta$ score
empirically  demonstrates that \textit{instruction level  explainability is  not
possible  at the  (C-)CFG level with GNNs, necessitating our novel application
of AFG and Meta-Coarsening approach}.

With respect to  \textbf{RQ3}, we find there are large  differences between both
malicious and benign classes. Specifically, we  observe that the benign class is
more  explainable than  predictable compared  to  the malicious  class and  vice
versa. Further  research is needed to  understand exactly why this  is the case,
though we provide some initial intuition in our work.

\subsubsection{Limitations} Lastly, several  limitations of our work  exist.
Specifically, it is  unclear if our method  will generalize to all other  types
of binary samples. However, this is extremely challenging to address without
additional datasets that contain malicious and benign \textit{binary executable
samples} specifically for Windows PE x86. Furthermore, some settings may contain
graphs with much larger samples than we use here. Thought, we do  provide a
framework  for working with  them by controlling for  $r$  and $\epsilon$.
Additionally, our coarsening approaches are largely deterministic and likely
much more expensive than state-of-the-art neural coarsening methods. However, to
the best of our knowledge, there are no easily implementable libraries available
for neural coarsening. If this were the case, then fast coarsening of  large
AFG graphs might  be feasible. While not part  of our work, our theoretical
\textit{end-to-end} pipeline necessitates a generation phase that in some cases
incurs a very large upfront dynamic generation cost in practice. Future work
could consider using statically generated samples CFGs or FCGs. Lastly, we do
not consider adversarial evasion techniques, zero days, or metamorphic code.
Adversarial training, and careful instruction perturbation might help towards
some of these issues.

\subsubsection{Ethical Considerations} In our work,  we primarily focus on
explainability for  the purpose of assisting defenders  to characterize
malware. However,  even though  our overall  goal is defensive in nature, it
could be  used to support adversaries. Theoretically, an adversary could used
our  work to evade detection of our  method via the removal or  perturbation  of
highly  characterized  malicious  subgraphs. Though  it  is unclear  as to  if
this  would  preserve the  overall malicious  nature of  such programs. While we
do not explore this  scenario, we expect that methods such as adversarial
training  could be used  as a potential mitigation.  Furthermore, we reasonably
expect that a much larger collection of data, specifically Windows PE x86
executable samples,  not generally  available to  the scientific  or public
communities, is needed to realize the evasion attack assisted by our method.

\section{Conclusion}
\label{sec:conclusion}

In our  work, we  propose a  novel approach,  Meta-Coarsening, that  can provide
granular assembly instruction level explanations for  malware detection using
GNNs.  To the best of our  knowledge, we are the first  to propose a GNN-based
technique that  is able  to learn  and explain  at varying levels of program
flow  down to  the  assembly level.  Most importantly,  our proposed approach is
capable of operating on large graphs that would otherwise be computationally
demanding in  practice. By applying coarsening-based methods, such  as  Kron
and  Variation Edges,  our  approach maintains  computationally efficiency
while  ensuring  performant  inference and  strong  explainability. Furthermore,
to  make our  approach  practical  in various  settings  different
hyperparameters  such as  $\epsilon$, $r$  can be used to  control the  size of
graphs.  Additionally, we  also  provide a  well defined explanation  selection
strategy, $\pi$  and $\tau$ to enhance explainability.

Our results show that at specific levels coarsening can improve  inference and
explainability. Furthermore,  our results  illustrate an important finding. That
benign samples tend to  be much more  explainable than malicious  samples in
our setting. This and  other explainability  results are supported  thought the
use  of the strongest  available  explanation  metric, characterization score,
that considers  both  sufficiency  and  necessity  of explanations not
considered in other works.

We conclude that our work has accomplished its goal of providing assembly level
explanations over AFG graphs  thought GNNs. It is our  hope that our work
serves  as inspiration  for others  to  evaluate large  scale malware  detection
explainability and can assist defenders against malware thought our approach.

\bibliographystyle{IEEEtran}
\bibliography{ref}

\begin{thebibliography}{10}
\providecommand{\url}[1]{#1}
\csname url@samestyle\endcsname
\providecommand{\newblock}{\relax}
\providecommand{\bibinfo}[2]{#2}
\providecommand{\BIBentrySTDinterwordspacing}{\spaceskip=0pt\relax}
\providecommand{\BIBentryALTinterwordstretchfactor}{4}
\providecommand{\BIBentryALTinterwordspacing}{\spaceskip=\fontdimen2\font plus
\BIBentryALTinterwordstretchfactor\fontdimen3\font minus \fontdimen4\font\relax}
\providecommand{\BIBforeignlanguage}[2]{{%
\expandafter\ifx\csname l@#1\endcsname\relax
\typeout{** WARNING: IEEEtran.bst: No hyphenation pattern has been}%
\typeout{** loaded for the language `#1'. Using the pattern for}%
\typeout{** the default language instead.}%
\else
\language=\csname l@#1\endcsname
\fi
#2}}
\providecommand{\BIBdecl}{\relax}
\BIBdecl

\bibitem{bilot_survey_2024}
\BIBentryALTinterwordspacing
T.~Bilot, N.~El~Madhoun, K.~Al~Agha, and A.~Zouaoui, ``A survey on malware detection with graph representation learning,'' vol.~56, no.~11, pp. 1--36. [Online]. Available: \url{https://dl.acm.org/doi/10.1145/3664649}
\BIBentrySTDinterwordspacing

\bibitem{amara_graphframex_2024}
\BIBentryALTinterwordspacing
K.~Amara, R.~Ying, Z.~Zhang, Z.~Han, Y.~Shan, U.~Brandes, S.~Schemm, and C.~Zhang, ``{GraphFramEx}: Towards systematic evaluation of explainability methods for graph neural networks,'' \emph{arXiv preprint}, no. {arXiv}:2206.09677, 2024. [Online]. Available: \url{http://arxiv.org/abs/2206.09677}
\BIBentrySTDinterwordspacing

\bibitem{anderson_graph-based_2011}
\BIBentryALTinterwordspacing
B.~Anderson, D.~Quist, J.~Neil, C.~Storlie, and T.~Lane, ``\BIBforeignlanguage{en}{Graph-based malware detection using dynamic analysis},'' \emph{\BIBforeignlanguage{en}{Journal in Computer Virology}}, vol.~7, no.~4, pp. 247--258, Nov. 2011. [Online]. Available: \url{https://doi.org/10.1007/s11416-011-0152-x}
\BIBentrySTDinterwordspacing

\bibitem{runwal_opcode_2012}
\BIBentryALTinterwordspacing
N.~Runwal, R.~M. Low, and M.~Stamp, ``\BIBforeignlanguage{en}{Opcode graph similarity and metamorphic detection},'' \emph{\BIBforeignlanguage{en}{Journal in Computer Virology}}, vol.~8, no.~1, pp. 37--52, May 2012. [Online]. Available: \url{https://doi.org/10.1007/s11416-012-0160-5}
\BIBentrySTDinterwordspacing

\bibitem{hashemi_graph_2017}
\BIBentryALTinterwordspacing
H.~Hashemi, A.~Azmoodeh, A.~Hamzeh, and S.~Hashemi, ``\BIBforeignlanguage{en}{Graph embedding as a new approach for unknown malware detection},'' \emph{\BIBforeignlanguage{en}{Journal of Computer Virology and Hacking Techniques}}, vol.~13, no.~3, pp. 153--166, Aug. 2017. [Online]. Available: \url{https://doi.org/10.1007/s11416-016-0278-y}
\BIBentrySTDinterwordspacing

\bibitem{khalilian_g3md_2018}
\BIBentryALTinterwordspacing
A.~Khalilian, A.~Nourazar, M.~Vahidi-Asl, and H.~Haghighi, ``G3md: Mining frequent opcode sub-graphs for metamorphic malware detection of existing families,'' vol. 112, pp. 15--33. [Online]. Available: \url{https://www.sciencedirect.com/science/article/pii/S0957417418303580}
\BIBentrySTDinterwordspacing

\bibitem{kakisim_metamorphic_2020}
\BIBentryALTinterwordspacing
A.~G. Kakisim, M.~Nar, and I.~Sogukpinar, ``Metamorphic malware identification using engine-specific patterns based on co-opcode graphs,'' \emph{Computer Standards \& Interfaces}, vol.~71, p. 103443, Aug. 2020. [Online]. Available: \url{https://www.sciencedirect.com/science/article/pii/S0920548919302685}
\BIBentrySTDinterwordspacing

\bibitem{fok_clustering_2021}
\BIBentryALTinterwordspacing
F.~K. Wai and V.~L.~L. Thing, ``{ Clustering based opcode graph generation for malware variant detection },'' in \emph{2021 18th International Conference on Privacy, Security and Trust (PST)}.\hskip 1em plus 0.5em minus 0.4em\relax Los Alamitos, CA, USA: IEEE Computer Society, Dec. 2021, pp. 1--11. [Online]. Available: \url{https://doi.ieeecomputersociety.org/10.1109/PST52912.2021.9647814}
\BIBentrySTDinterwordspacing

\bibitem{kakisim_sequential_2022}
\BIBentryALTinterwordspacing
A.~G. Kakisim, S.~Gulmez, and I.~Sogukpinar, ``Sequential opcode embedding-based malware detection method,'' \emph{Computers \& Electrical Engineering}, vol.~98, p. 107703, 2022. [Online]. Available: \url{https://www.sciencedirect.com/science/article/pii/S0045790622000210}
\BIBentrySTDinterwordspacing

\bibitem{shokouhinejad_consistency_2025}
\BIBentryALTinterwordspacing
H.~Shokouhinejad, G.~Higgins, R.~Razavi-Far, H.~Mohammadian, and A.~A. Ghorbani, ``On the consistency of gnn explanations for malware detection,'' \emph{Information Sciences}, vol. 721, p. 122603, 2025. [Online]. Available: \url{https://www.sciencedirect.com/science/article/pii/S0020025525007364}
\BIBentrySTDinterwordspacing

\bibitem{herath_cfgexplainer_2022}
\BIBentryALTinterwordspacing
J.~D. Herath, P.~P. Wakodikar, P.~Yang, and G.~Yan, ``{CFGExplainer}: Explaining graph neural network-based malware classification from control flow graphs,'' in \emph{2022 52nd Annual {IEEE}/{IFIP} International Conference on Dependable Systems and Networks ({DSN})}, pp. 172--184, {ISSN}: 2158-3927. [Online]. Available: \url{https://ieeexplore.ieee.org/document/9833560}
\BIBentrySTDinterwordspacing

\bibitem{shokouhinejad_dual_2025}
\BIBentryALTinterwordspacing
H.~Shokouhinejad, R.~Razavi-Far, G.~Higgins, and A.~A. Ghorbani, ``Dual explanations via subgraph matching for malware detection,'' \emph{arxiv prerprint}, no. {arXiv}:2504.20904. [Online]. Available: \url{http://arxiv.org/abs/2504.20904}
\BIBentrySTDinterwordspacing

\bibitem{shokouhinejad_recent_2025}
\BIBentryALTinterwordspacing
H.~Shokouhinejad, R.~Razavi-Far, H.~Mohammadian, M.~Rabbani, S.~Ansong, G.~Higgins, and A.~A. Ghorbani, ``Recent advances in malware detection: Graph learning and explainability,'' \emph{arXiv preprint}, no. {arXiv}:2502.10556, 2025. [Online]. Available: \url{http://arxiv.org/abs/2502.10556}
\BIBentrySTDinterwordspacing

\bibitem{karl-bridge-microsoft_pe_nodate}
\BIBentryALTinterwordspacing
{Karl-Bridge-Microsoft}, ``\BIBforeignlanguage{en-us}{{PE} {Format} - {Win32} apps}.'' [Online]. Available: \url{https://learn.microsoft.com/en-us/windows/win32/debug/pe-format}
\BIBentrySTDinterwordspacing

\bibitem{shoshitaishvili2016state}
Y.~Shoshitaishvili, R.~Wang, C.~Salls, N.~Stephens, M.~Polino, A.~Dutcher, J.~Grosen, S.~Feng, C.~Hauser, C.~Kruegel, and G.~Vigna, ``Sok: (state of) the art of war: Offensive techniques in binary analysis,'' in \emph{IEEE Symposium on Security and Privacy}, 2016.

\bibitem{stephens2016driller}
N.~Stephens, J.~Grosen, C.~Salls, A.~Dutcher, R.~Wang, J.~Corbetta, Y.~Shoshitaishvili, C.~Kruegel, and G.~Vigna, ``Driller: Augmenting fuzzing through selective symbolic execution,'' in \emph{NDSS}, 2016.

\bibitem{shoshitaishvili2015firmalice}
Y.~Shoshitaishvili, R.~Wang, C.~Hauser, C.~Kruegel, and G.~Vigna, ``Firmalice - automatic detection of authentication bypass vulnerabilities in binary firmware,'' in \emph{NDSS}, 2015.

\bibitem{loukas_graph_2018}
\BIBentryALTinterwordspacing
A.~Loukas, ``Graph reduction with spectral and cut guarantees,'' \emph{Journal of Machine Learning Research}, vol.~20, no. 116, pp. 1--42, 2019. [Online]. Available: \url{http://jmlr.org/papers/v20/18-680.html}
\BIBentrySTDinterwordspacing

\bibitem{qemu_capstoneincludecapstonex86h_nodate}
{qemu}, ``\BIBforeignlanguage{en}{capstone/include/capstone/x86.h at master · qemu/capstone},'' \url{https://github.com/qemu/capstone/blob/master/include/capstone/x86.h}.

\bibitem{dikedataset}
G.-A. Iosif, ``Dikedataset,'' \url{https://github.com/iosifache/DikeDataset}, 2021, accessed on February 27, 2024.

\bibitem{practicalsecurity2024pe}
{Practical Security Analytics LLC}, ``Pe malware machine learning dataset,'' \url{https://practicalsecurityanalytics.com/pe-malware-machine-learning-dataset/}, 2024, accessed: 2024-08-06.

\bibitem{kipf_semi-supervised_2017}
\BIBentryALTinterwordspacing
T.~N. Kipf and M.~Welling, ``Semi-supervised classification with graph convolutional networks,'' \emph{arXiv preprint}, no. {arXiv}:1609.02907, 2017. [Online]. Available: \url{http://arxiv.org/abs/1609.02907}
\BIBentrySTDinterwordspacing

\bibitem{Captum}
\BIBentryALTinterwordspacing
N.~Kokhlikyan, V.~Miglani, M.~Martin, E.~Wang, B.~Alsallakh, J.~Reynolds, A.~Melnikov, N.~Kliushkina, C.~Araya, S.~Yan, and O.~Reblitz-Richardson, ``Captum: A unified and generic model interpretability library for pytorch,'' \emph{arXiv preprint}, no. {arXiv}:2009.07896, 2020. [Online]. Available: \url{https://arxiv.org/abs/2009.07896}
\BIBentrySTDinterwordspacing

\bibitem{GNNExplainer}
Z.~Ying, D.~Bourgeois, J.~You, M.~Zitnik, and J.~Leskovec, ``Gnnexplainer: Generating explanations for graph neural networks,'' \emph{Advances in Neural Information Processing Systems (NIPS)}, vol.~32, 2019.

\bibitem{PGExplainer}
D.~Luo, W.~Cheng, D.~Xu, W.~Yu, B.~Zong, H.~Chen, and X.~Zhang, ``Parameterized explainer for graph neural network,'' in \emph{34th International Conference on Neural Information Processing Systems}, 2020.

\bibitem{shuman_multiscale_2016}
\BIBentryALTinterwordspacing
D.~I. Shuman, M.~J. Faraji, and P.~Vandergheynst, ``A {Multiscale} {Pyramid} {Transform} for {Graph} {Signals},'' \emph{IEEE Transactions on Signal Processing}, vol.~64, no.~8, pp. 2119--2134, Apr. 2016. [Online]. Available: \url{https://ieeexplore.ieee.org/document/7366599/}
\BIBentrySTDinterwordspacing

\bibitem{Fey_Lenssen_2019}
M.~Fey and J.~E. Lenssen, ``Fast graph representation learning with {PyTorch Geometric},'' in \emph{ICLR Workshop on Representation Learning on Graphs and Manifolds}, 2019.

\bibitem{Fey_etal_2025}
M.~Fey, J.~Sunil, A.~Nitta, R.~Puri, M.~Shah, B.~Stojanovi{\v{c}}, R.~Bendias, B.~Alexandria, V.~Kocijan, Z.~Zhang, X.~He, J.~E. Lenssen, and J.~Leskovec, ``{PyG} 2.0: Scalable learning on real world graphs,'' in \emph{Temporal Graph Learning Workshop @ KDD}, 2025.

\bibitem{scipyproceedings_11}
A.~A. Hagberg, D.~A. Schult, and P.~J. Swart, ``Exploring network structure, dynamics, and function using networkx,'' in \emph{Proceedings of the 7th Python in Science Conference}, G.~Varoquaux, T.~Vaught, and J.~Millman, Eds., Pasadena, CA USA, 2008, pp. 11 -- 15.

\end{thebibliography}

\appendix
\subsection{Explainer Comparison and Characterization Curve Analysis}
\label{charact_analysis}

Here, we provide additional explainability analysis and comparison of IG with
Guided Back Propogation (GBP) and Salience (SAL) explainers from the Captum
library \cite{Captum}. Specifically, at each explanation level we compute the
impact of sparsity on characterization accuracy. However, at the AFG level all
explanations are derived from IG output to form a consistent comparison.

\begin{figure*}[t]
    \centering
    \begin{minipage}{0.23\textwidth}
        \centering
        \subfloat[CFG ($r=0$)]{\includegraphics[width=\textwidth]{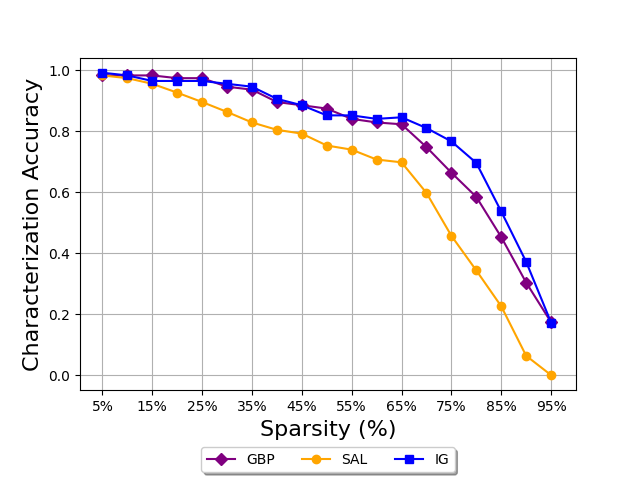}}
    \end{minipage}
    \hfil
    \begin{minipage}{0.23\textwidth}
        \centering
        \subfloat[AFG ($r=0$)]{\includegraphics[width=\textwidth]{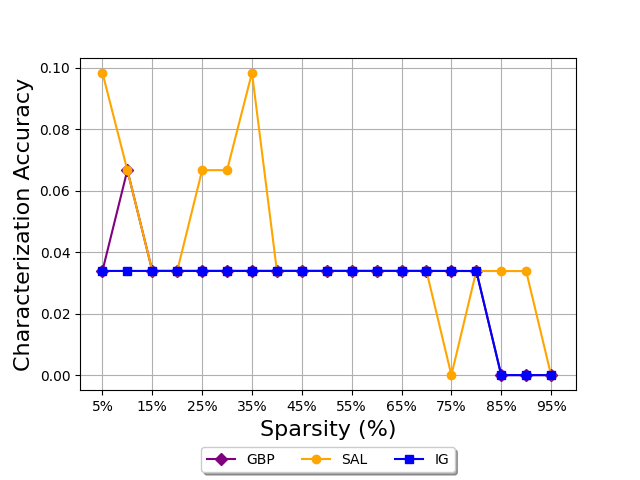}}
    \end{minipage}

    \caption{Characterization accuracy curves for \textbf{Baseline}, i.e., no coarsening, with \textbf{$r=0$}.}
    \label{fig:baseline_charact}
\end{figure*}

\begin{figure*}[t]
    \centering
    \begin{minipage}{0.23\textwidth}
        \centering
        \subfloat[C-CFG ($r=0.25$)]{\includegraphics[width=\textwidth]{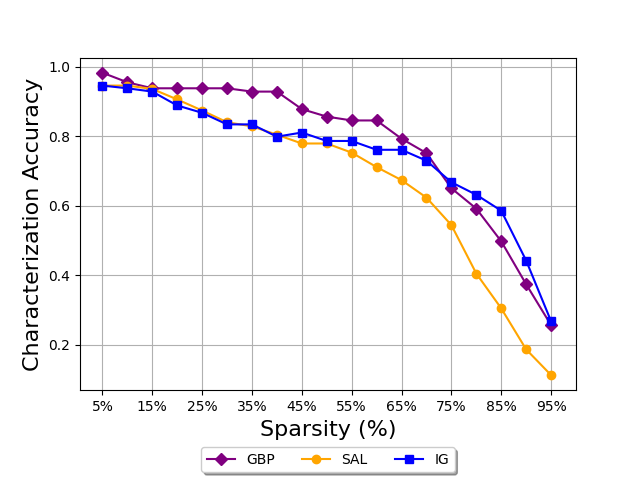}} \\
        \subfloat[AFG ($r=0.25$)]{\includegraphics[width=\textwidth]{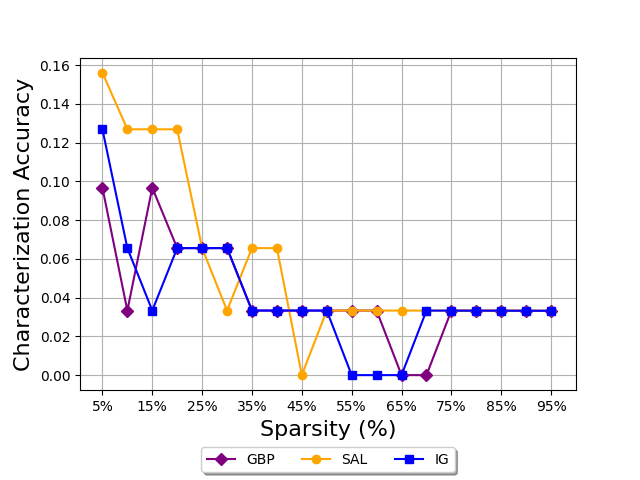}}
    \end{minipage}
    \hfill
    \begin{minipage}{0.23\textwidth}
        \centering
        \subfloat[C-CFG ($r=0.50$)]{\includegraphics[width=\textwidth]{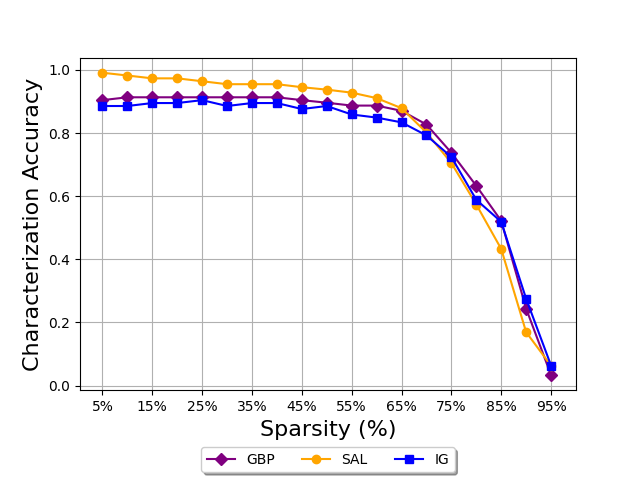}} \\
        \subfloat[AFG ($r=0.50$)]{\includegraphics[width=\textwidth]{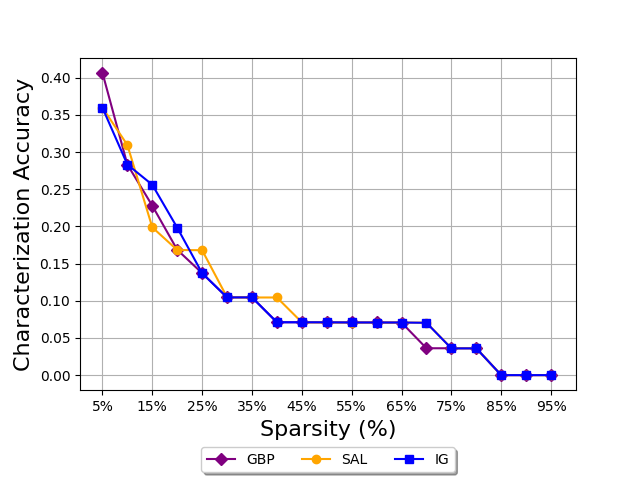}}
    \end{minipage}
    \hfill
    \begin{minipage}{0.23\textwidth}
        \centering
        \subfloat[C-CFG ($r=0.75$)]{\includegraphics[width=\textwidth]{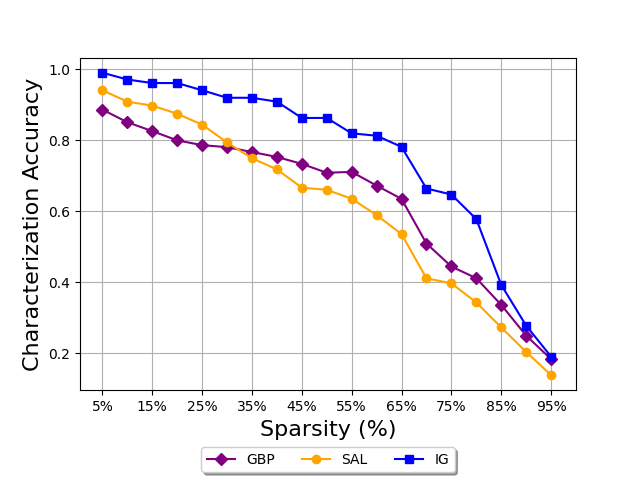}} \\
        \subfloat[AFG ($r=0.75$)]{\includegraphics[width=\textwidth]{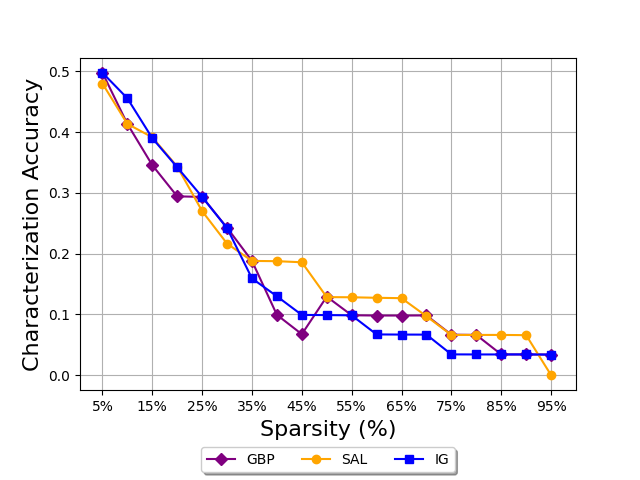}}
    \end{minipage}
    \hfill
    \begin{minipage}{0.23\textwidth}
        \centering
        \subfloat[C-CFG ($r=0.999$)]{\includegraphics[width=\textwidth]{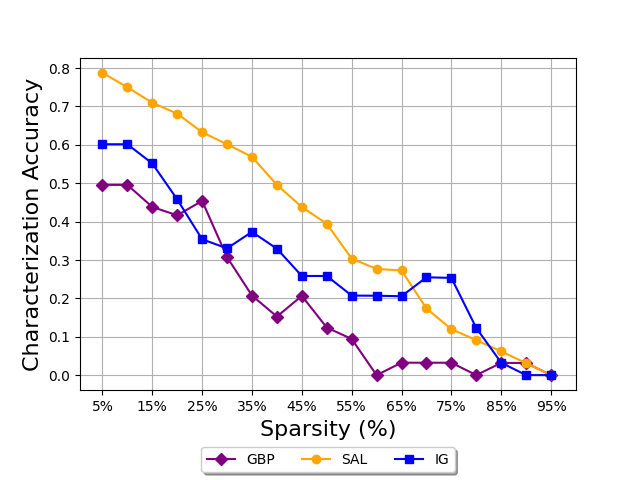}} \\
        \subfloat[AFG ($r=0.999$)]{\includegraphics[width=\textwidth]{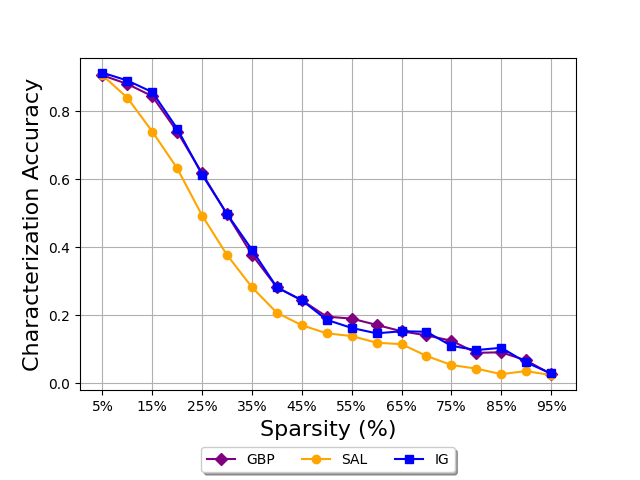}}
    \end{minipage}

    \caption{Characterization accuracy curves for \textbf{Variation Edges} with \textbf{$r \in \{0.25, 0.5, 0.75, 0.999\}$}.}
    \label{fig:ve_charact}
\end{figure*}

\begin{figure*}[t]
    \centering
    \begin{minipage}{0.23\textwidth}
        \centering
        \subfloat[C-CFG ($r=0.25$)]{\includegraphics[width=\textwidth]{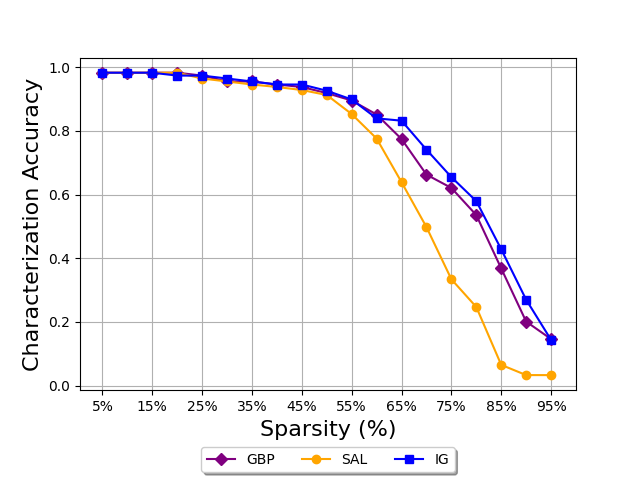}} \\
        \subfloat[AFG ($r=0.25$)]{\includegraphics[width=\textwidth]{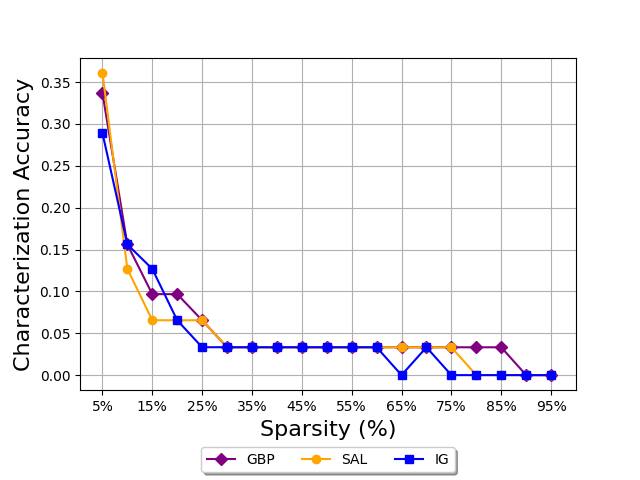}}
    \end{minipage}
    \hfill
    \begin{minipage}{0.23\textwidth}
        \centering
        \subfloat[C-CFG ($r=0.50$)]{\includegraphics[width=\textwidth]{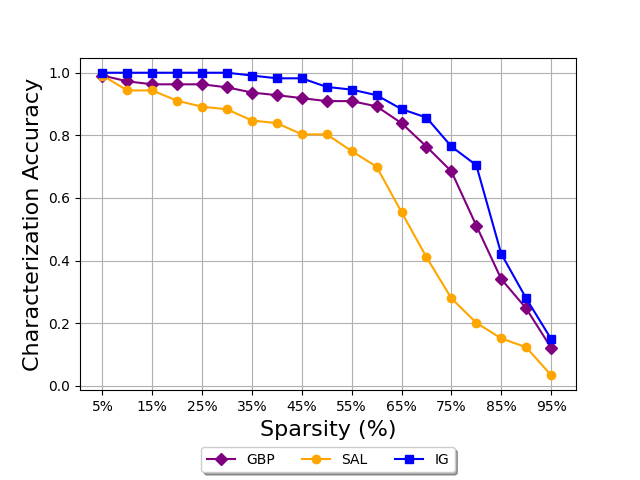}} \\
        \subfloat[AFG ($r=0.50$)]{\includegraphics[width=\textwidth]{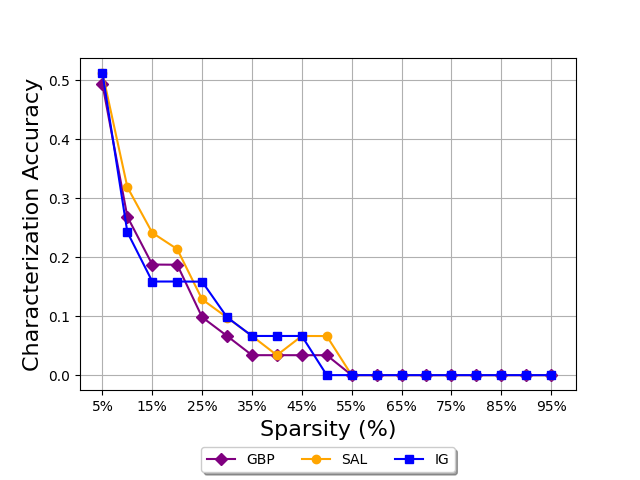}}
    \end{minipage}
    \hfill
    \begin{minipage}{0.23\textwidth}
        \centering
        \subfloat[C-CFG ($r=0.75$)]{\includegraphics[width=\textwidth]{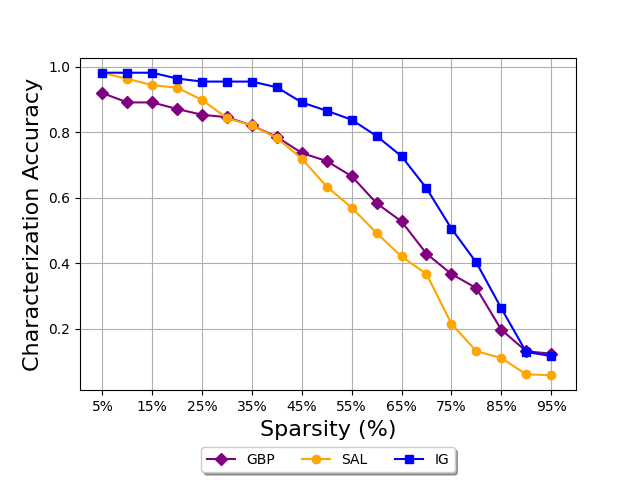}} \\
        \subfloat[AFG ($r=0.75$)]{\includegraphics[width=\textwidth]{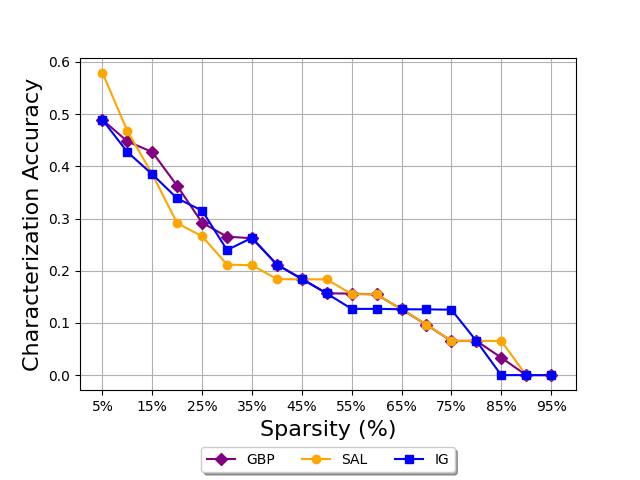}}
    \end{minipage}
    \hfill
    \begin{minipage}{0.23\textwidth}
        \centering
        \subfloat[C-CFG ($r=0.999$)]{\includegraphics[width=\textwidth]{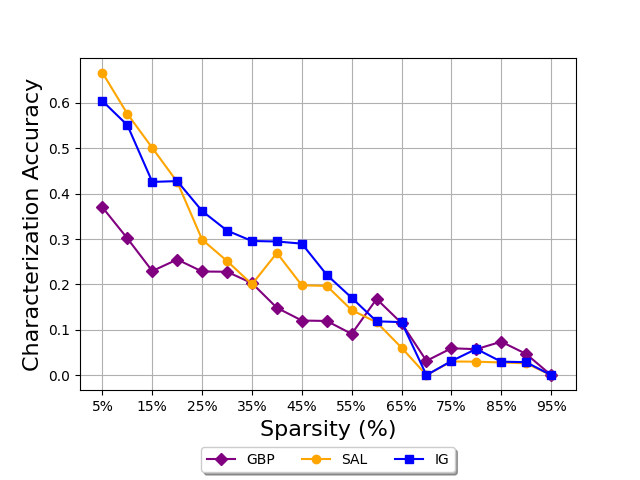}} \\
        \subfloat[AFG ($r=0.999$)]{\includegraphics[width=\textwidth]{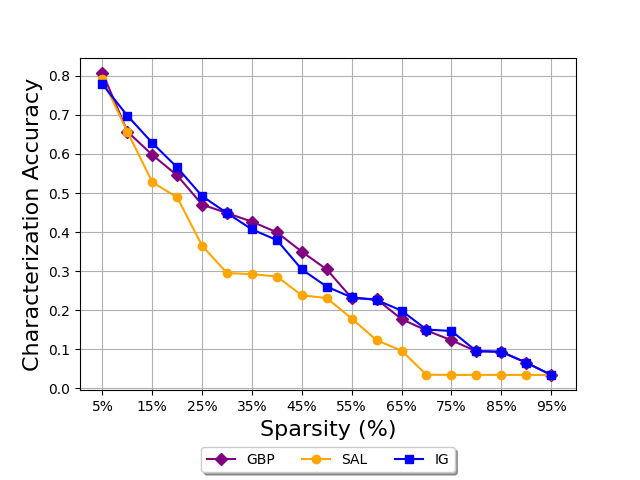}}
    \end{minipage}

    \caption{Characterization accuracy curves for \textbf{Kron} with \textbf{$r \in \{0.25, 0.5, 0.75, 0.999\}$}.}
    \label{fig:kron_charact}
\end{figure*}

When examining general  explanation trends over all experiments,  we observe
that in many  cases at the  (C-)CFG  level,  shown in  Figures
\ref{fig:baseline_charact} panel \textit{a}, \ref{fig:ve_charact} panel
\textit{e}, and \ref{fig:kron_charact} panels \textit{a}, \textit{c}, \textit{e}
that at  IG,  GBP, and  SAL perform best, from high to low,  respectively.
However,  in  some  cases  when using Variation Edges, SAL  tends  to outperform
other explainers,  shown  in Figure \ref{fig:ve_charact} panels \textit{c} and
\textit{g}. Furthermore, at the B-AFG  level we  generally observe similar
performance across explainers, shown  in Figures \ref{fig:baseline_charact} panel \textit{b}, \ref{fig:ve_charact} panel \textit{b}, \ref{fig:kron_charact} panel \textit{b}, \textit{d}, \textit{f}, and \ref{fig:kron_charact} \textit{b}, \textit{d}, \textit{f}. However, in  some cases we also observe the previous trend of  IG, GBP, and SAL from  high to low, respectively, shown in Figures \ref{fig:ve_charact} panel \ref{fig:kron_charact} all panels \textit{h}, when coarsening  is maximal. 

\end{document}